\documentclass[%
 reprint,
 prx,
 floatfix,
 amsmath,amssymb, 
 aps, widetext, superscriptaddress,nofootinbib
]{revtex4-2}

\usepackage{upgreek}
\usepackage{orcidlink}
\usepackage{xcolor}
\usepackage{hyperref}
\usepackage{graphicx}
\usepackage{dcolumn}
\usepackage{bm}
\usepackage{microtype}
\usepackage{physics}
\usepackage{xspace}
\usepackage{chemformula}
\usepackage{csquotes}
\usepackage{upgreek}
\usepackage{lipsum}
\usepackage{siunitx}
\usepackage{todonotes}
\usepackage[english]{babel}

\newcommand{\wq}{\omega_\mathrm{q}}
\newcommand{\kq}{k_\mathrm{q}}
\newcommand{\vg}{v_\mathrm{g}}
\newcommand{\laplace}[1]{\tilde{#1}}

\allowdisplaybreaks[3]

\newcommand{\tuwien}{Institute for Theoretical Physics and Vienna Center for Quantum Science and Technology, Vienna University of Technology (TU Wien), Vienna A-1040, Austria}
\newcommand{\palermo}{Università degli Studi di Palermo, Dipartimento di Fisica e Chimica-Emilio Segrè, Via Archirafi 36, 90123 Palermo, Italy}

\newcommand{\hll}[1]{{#1}}

\begin{document}
\title{Enabling Deterministic Passive Quantum State Transfer with Giant Atoms}

\author{Oliver Diekmann\,\orcidlink{0000-0001-7920-3641}}
\affiliation{\tuwien}

\author{Enrico Di Benedetto\,\orcidlink{0009-0003-3613-5257}}
\affiliation{\palermo}

\author{Nicolas Jungwirth\,\orcidlink{0009-0001-7826-1791}}
\altaffiliation{Current address: Institut für Experimentalphysik, Universität Innsbruck, Technikerstraße 25/4, 6020 Innsbruck, Austria}
\affiliation{TUM School of Natural Sciences, Technical University of Munich, D-85748 Garching, Germany}
\affiliation{Walther-Meißner-Institut, Bayerische Akademie der Wissenschaften, D-85748 Garching, Germany}
\affiliation{Munich Center for Quantum Science and Technology (MCQST), D-80799 Munich, Germany}

\author{Daniele De Bernardis\,\orcidlink{0000-0002-9618-0389}}
\affiliation{Istituto Nazionale di Ottica (CNR-INO), c/o LENS via Nello Carrara 1, Sesto F.no 50019, Italy}

\author{Zeyu Kuang\,\orcidlink{0000-0001-7258-6643}}
\affiliation{\tuwien}

\author{Francesco Ciccarello\,\orcidlink{0000-0002-6061-1255}}
\affiliation{\palermo}
\affiliation{NEST, Istituto Nanoscienze-CNR, Piazza S. Silvestro 12, 56127 Pisa (Italy)}

\author{Stefan Rotter\,\orcidlink{0000-0002-4123-1417}}
\affiliation{\tuwien}

\author{Peter Rabl\,\orcidlink{0000-0002-2560-8835}}
\affiliation{TUM School of Natural Sciences, Technical University of Munich, D-85748 Garching, Germany}
\affiliation{Walther-Meißner-Institut, Bayerische Akademie der Wissenschaften, D-85748 Garching, Germany}
\affiliation{Munich Center for Quantum Science and Technology (MCQST), D-80799 Munich, Germany}

\author{Alejandro González-Tudela\,\orcidlink{0000-0003-2307-6967}}
\email{a.gonzalez.tudela@csic.es}
\affiliation{Quantum Advanced Research Center (QuARC), CSIC, Calle Serrano 113b, 28006 Madrid, Spain}
\affiliation{Institute of Fundamental Physics (IFF), CSIC, Calle Serrano 113b, 28006 Madrid, Spain}
\author{Carlos Gonzalez-Ballestero\,\orcidlink{0000-0002-7639-0856}}
\email{carlos.gonzalez-ballestero@tuwien.ac.at}
\affiliation{\tuwien}

\date{\today}

\begin{abstract}
Achieving quantum state transfer in passive ways can become a powerful asset for scalable quantum networks.
Here, we demonstrate how giant atoms coupled to 1D waveguides provide a platform for such a passive, deterministic transfer. 
\hll{Specifically,} we show that \hll{when the giant atom's extent is comparable to the width of the emitted wavepacket, this wavepacket can be designed to be} 
time-reversal-symmetric \hll{by engineering the positions and strengths of atom-waveguide coupling points.} 
We first derive general analytical conditions under which arbitrary qubit decays can be mapped to wavevector-dependent couplings that guarantee perfect state transfer in the continuum limit of infinitely many coupling points. Then, for experimentally relevant configurations with a finite number of coupling points, we \hll{optimize the coupling positions and values. We }demonstrate  that \hll{the transfer fidelity can be boosted well beyond the value for a single coupling point (54\,\%),}
reaching 87\% with only two coupling points and exceeding 99\% with ten or more. We further analyze the robustness of the protocol against  
\hll{disorder in the coupling points as well as frequency variations of the qubits}
and extend the formalism to environments with nonlinear dispersion, showing that dispersion-induced distortions can be fully compensated by judiciously chosen setups. Our results establish giant atoms as a powerful platform for realizing high-fidelity quantum state transfer \hll{and demonstrate how time-dependent control can be instead encoded into the device design.}
\end{abstract}

\maketitle
Artificial atoms coupled to engineered photonic environments are at the forefront of developments in quantum science~\cite{lodahl_interfacing_2015,krantz_quantum_2019,bradac_quantum_2019}. One particular paradigm of quantum emitters that has attracted increasing interest over the past decade is the concept of \textit{giant atoms}~\cite{frisk_kockum_quantum_2021}, i.e., quantum emitters that are coupled to their environment at more than a single spatial point. Such non-local coupling has been achieved with surface acoustic waves~\cite{gustafsson_propagating_2014,andersson_non-exponential_2019}, microwave transmission lines~\cite{kannan_waveguide_2020, vadiraj_engineering_2021}, magnonics~\cite{wang_giant_2022} as well as coupled cavity arrays~\cite{jouanny_superstrong_2025}, and extensions to other platforms have been proposed~\cite{gonzalez-tudela_engineering_2019,chen_giant-atom_2023}. The self-interference of the emission from different \textit{legs} gives rise to intricate behavior such as strongly frequency-dependent decay rates~\cite{frisk_kockum_designing_2014, yu_entanglement_2021}, decoherence-free interactions~\cite{kockum_decoherence-free_2018, kannan_waveguide_2020,carollo_mechanism_2020, soro_chiral_2022, leonforte_quantum_2025, wang_decoherence-free_2026}, unconventional bound states~\cite{guo_oscillating_2020,vega_qubit-photon_2021,wang_tunable_2021, xiao_bound_2022, lim_oscillating_2023} and non-Markovian dynamics~\cite{guo_giant_2017,guo_beyond_2020,andersson_non-exponential_2019,longhi_photonic_2020, du_single-photon_2021, yin_non-markovian_2022,qiu_collective_2023,wang_realizing_2024,roccati_controlling_2024,gonzalez-gutierrez_non-markovian_2025}, allowing for applications, e.g., in open quantum system simulations~\cite{chen_simulating_2025} or quantum computation~\cite{levy-yeyati_passive_2025,levy-yeyati_engineering_2026,chen_efficient_2026, rieck_controlled-z_2026}.
\begin{figure}
    \centering
    \includegraphics[]{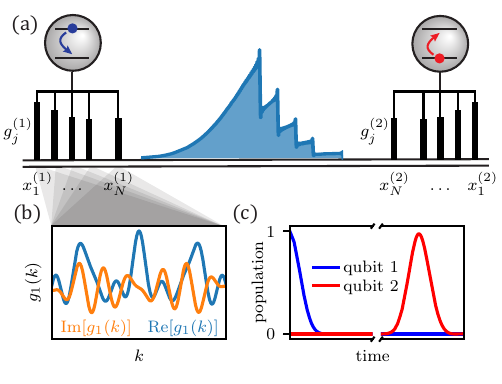}
    \caption{Passive quantum state transfer between giant atoms chirally coupled to a waveguide with a linear dispersion. (a) Two giant atoms are coupled to the waveguide at $N$ discrete coupling points $ x^{(\alpha)}_j, g^{(\alpha)}_j~(j=1,...,N;~\alpha=1,2)$ (see text for details). The nonlocal coupling allows the left qubit to emit an almost time-reversal-symmetric pulse (blue), that can be absorbed by the right qubit. (b) The interference of contributions from different coupling points gives rise to a wavevector-dependent coupling $g_1(k)$. (c) Resulting evolution of the excited-state populations $|c_\alpha(t)|^2$ of the two qubits. The configuration assumed in this figure is the result of the optimization in Fig.~\ref{fig:fig3} for $N=5$.}
    \label{fig:1}
\end{figure}

In this paper, we propose a novel application that directly leverages giant atom implementations: \textit{deterministic passive quantum state transfer}.
In their seminal work Ref.~\cite{cirac_quantum_1997}, Cirac {\it et al.}~first demonstrated how quantum states can be deterministically transferred between two distant nodes of a quantum network~\cite{kimble_quantum_2008} via the controlled emission and reabsorption of single photons. The solution developed in this work relies on time-dependent qubit-waveguide couplings to shape time-reversal-symmetric photon wavepackets, which are then perfectly reabsorbed at the second node using a time-reversed control pulse. 
This active approach has been successfully implemented in the microwave \cite{pechal_microwave-controlled_2014,kurpiers_deterministic_2018,axline_-demand_2018, magnard_microwave_2020,almanakly_deterministic_2025, miyamura_generation_2025, hernandez-anton_emission_2026} and optical domain~\cite{ritter_elementary_2012} as well as using surface acoustic waves~\cite{bienfait_phonon-mediated_2019, dumur_quantum_2021}, and many theoretical extensions have been proposed~\cite{vogell_deterministic_2017, morin_deterministic_2019,penas_improving_2023,penas_multiplexed_2024,he_quantum_2025,pernas_shaping_2026,jiang_photon-echo_2026,wang_chiral_2022,xu_catch_2024, du_dressed_2025,luo_dynamic_2025}.
\hll{As an alternative path to state transfer,} passive schemes have been investigated, for example, in spin chains~\cite{bose_quantum_2003} (see also~\cite{banchi_ballistic_2013, bernard_distinctive_2024} and references therein), photonic quantum Hall systems~\cite{de_bernardis_chiral_2023} and exotic cavities~\cite{diekmann_ultrafast_2024}, or by engineered cavity-cluster quantum nodes~\cite{tian_static_2021} and waveguide dispersions~\cite{kuang_perfect_2025}. 
In these 
passive state transfer schemes, a time-reversed waveform of the emitted photon is achieved by locally coupling a qubit to a complex photonic reservoir, which limits the practicality and scalability of this approach.

Here, we show that giant atoms offer a distinct approach to encode the required temporal structure directly into the spatial design of the quantum emitter. Specifically, by tuning their leg positions and couplings, it is possible to achieve passive state transfer between such giant emitters. 
As a major conceptual result, we first analytically prove that in the limit of infinitely many coupling points (i.e., for a continuous coupling distribution) there always exists a configuration yielding a time-reversal-symmetric pulse for any possible qubit decay into a linear-dispersion waveguide. We derive the associated wavevector-dependent coupling distributions and demonstrate the state transfer in the particular cases of a Gaussian and exponential decay. \hll{These examples demonstrate that deterministic passive state transfer is connected to the non-Markovian regime of giant atom physics, where the extent of the atoms is comparable to the width of the emitted pulse.} We then focus on a finite number of giant atom legs and show that the pulses mediating the state transfer can remain approximately time-reversal-symmetric provided that the coupling strengths and positions are properly engineered. Optimizing for the state transfer fidelity as a function of the number of legs of the giant atom, we find that already with two legs, we can  achieve a transfer probability of around $87\,\%$, which is readily increased beyond $99\,\%$ by employing ten or more legs. Lastly, we show that our proposal can be implemented in existing experimental platforms\hll{, which support the operation in the non-Markovian regime, for example surface acoustic waves or coupled-cavity arrays interacting with superconducting qubits}. In particular, we study the effect of imperfections in the leg positions\hll{, coupling strengths, and deviations in the qubit frequency,} and show how our theory can be extended to photonic baths with curved dispersion such as coupled cavity arrays.

This paper is organized as follows: In Sec.~\ref{sec:model} we introduce the system under consideration. In Sec.~\ref{sec:infiniteLegs} we study state transfer for a continuous spatial coupling distribution, i.e., the limit of infinitely many legs. In Sec.~\ref{sec:finiteLeg} we optimize the state transfer for a finite number of legs. Finally, in Sec.~\ref{sec:IV} we discuss experimental realizations and the robustness of our proposal, and conclude in Sec.~\ref{sec:conclusion}. 

\section{Model}\label{sec:model}
We consider two giant atoms coupled to a one-dimensional waveguide. The model Hamiltonian for this system is ($\hbar=1$)
\begin{align}
    \mathcal{H} &= \wq\sum_{\alpha=1,2}\sigma_\alpha^+\sigma_\alpha^-+\int_{-\infty}^{+\infty}\dd{k}\omega(k) a_k^\dagger a^{\phantom{\dagger}}_k\notag\\ &+\sum_{\alpha=1,2}\int_{-\infty}^{+\infty}\dd{k}(g_\alpha(k)\sigma^+_\alpha a^{\phantom{\dagger}}_k+\mathrm{H.c.})\,,
    \label{eq:Hamiltonian}
\end{align}
where we assume a linear dispersion $\omega(k)=\vg |k|$ with group velocity $\vg$. In Eq.~\eqref{eq:Hamiltonian}, $\sigma_\alpha^\pm$ are standard Pauli operators, $a^{\phantom{\dagger}}_k$ ($a_k^\dagger$) is the annihilation (creation) operator for photons at wavevector $k$ and $g_\alpha(k)$ is the wavevector-dependent coupling of qubit $\alpha$. 
Since the $\alpha$th qubit couples to the waveguide at several points $x_j^{(\alpha)}$ the coupling takes the form 
\begin{equation}
    g_\alpha(k) = \sum_je^{ikx^{(\alpha)}_j}g^{(\alpha)}_j, \label{eq:giantAtomCoupling}
\end{equation}
with the local coupling strengths $g^{(\alpha)}_j$ for the $j$th leg of the $\alpha$th qubit. \hll{We take the qubits to be arranged in a serial configuration such that $x^{(1)}_j<x^{(2)}_k$ for any $j,k$ and assume them to decay on a timescale $\tau\gg\wq^{-1}$ (for all numerical calculations we will assume $\tau=10^4 \wq^{-1}/\pi $).} \hll{Note that the conventional $\sqrt{|k|}$ dependence of $g(k)$ can be approximated as constant, as it changes negligibly on the relevant frequency scale $\tau^{-1}$ around the qubit resonance frequency $\wq$.} Note that for an atom coupled to a waveguide at a single point (i.e., a \textit{normal} atom), $g_\alpha(k)$ is typically considered a constant function of $k$. 
The effect of multiple coupling points (\textit{giant} atom) is to introduce a strong $k$-dependence of the coupling function due to interference of multiple coupling points.
In the following we will base our derivations and optimizations on the assumption of a chiral coupling to the waveguide, i.e., the qubits only couple to right-propagating modes. While this allows us to focus on configurations such as the one shown in Fig.~\ref{fig:1}, it is not a fundamental restriction of our work as we provide explicit ways to achieve this chiral coupling for continuous (cf.~Sec.~\ref{sec:infinite:examples}) and discrete (see Appendix~\ref{sec:appendixChirality}) coupling configurations.

Throughout this work, we focus on the dynamics in the single excitation sector, where the state of the system can be written in the general form
\begin{align}
    \ket{\psi(t)} &= c_1(t)e^{-i\wq t}\ket{e_1}+c_2(t)e^{-i\wq t}\ket{e_2}\notag\\ &+\int_{-\infty}^{+\infty}\dd{k}c_k(t)e^{-i\omega(k) t}\ket{1_k}\,.
    \label{eq:wavefunction}
\end{align}
Here, $\ket{e_1}\equiv\ket{e}\ket{g}\ket{\mathrm{vac}}$ ($\ket{e_2}\equiv\ket{g}\ket{e}\ket{\mathrm{vac}}$) denotes the state where only the first (second) qubit is excited and $\ket{1_k}\equiv a_k^{\dag}\ket{g}\ket{g}\ket{\mathrm{vac}}$ the state where only the $k$th mode is occupied by a single photon. 
Plugging this ansatz into the Schr\"odinger equation,
we obtain a set of coupled equations of motions for the coefficients of Eq.~\eqref{eq:wavefunction},
\begin{align}
    \dot{{c}}_{1(2)}(t) &= -i\int_{-\infty}^{+\infty} \dd{k}g_{1(2)}(k) c_k(t)e^{-i\Delta(k) t}\,,\notag\\
    \dot{{c}}_k(t) &= -i\left[g_1^*(k) c_1(t)+g_2^*(k) c_2(t)\right]e^{i\Delta(k) t}\,,
    \label{eq:eom}
\end{align}
with the detuning $\Delta(k) = \omega(k)-\wq$.

In the following, we will study conditions for the perfect transfer of an excitation from the first qubit (initially in its excited state) to the second one (initially in the ground state).
Note that the successful transfer of an excitation also entails the transfer of arbitrary superposition states of the first qubit \cite{cirac_quantum_1997}. 

\section{State transfer between qubits with a continuous spatial coupling}
\label{sec:infiniteLegs}
As a fundamental paradigm, we consider the limiting case of a continuous coupling distribution~\cite{guo_oscillating_2020,lin_giant_2026}, which can be thought of as giant atoms with infinitely many legs. We explicitly assume a coupling to right-propagating modes only and relax this limitation later on.
In this case, Eq.~\eqref{eq:giantAtomCoupling} approaches the Fourier transform of a continuous spatial coupling distribution $g_\alpha(x)$,
\begin{equation}
    g_\alpha(k)=\int_{-\infty}^{+\infty}\dd{x}e^{ikx}g_\alpha(x)\,.
    \label{eq:spatialCouplings}
\end{equation}
In the following, our goal is to derive wavevector-dependent couplings $g_1(k)$ that mediate perfect state transfer with time-reversal-symmetric pulses. These couplings can then be converted to spatial distributions $g_1(x)$ by inverting the above equation.

The serial configuration of the giant atoms, together with the assumption of chiral emission into the waveguide, implies that we can consider the emission of the first qubit and the absorption by the second one separately. For this reason, we first focus on the dynamics of the first qubit only.
With this in mind, the exact evolution of the first qubit amplitude can be computed by the resolvent operator approach~\cite{cohen-tannoudji_atom-photon_2008, gonzalez-tudela_markovian_2017},
\begin{align}
    \label{eq:resolvent-main}
    c_1(t) = \frac{i}{2\pi}\int_{-\infty}^{\infty}\dd{\Delta}\frac{e^{-i\Delta t}}{\Delta-\Sigma(\Delta+\wq+i0^+)}\,,
\end{align}
where the self-energy $\Sigma(E)$ is given by
\begin{equation}
    \Sigma(E) = \int_0^{+\infty}\dd{k}\frac{|g_1(k)|^2}{{E}-\omega(k)}\,.
\end{equation}
Typically, given a certain coupling configuration $g_1(k)$, we can solve for the exact dynamics through Eq.~\eqref{eq:resolvent-main}. In the present context, though, it is more meaningful to ask whether we can solve for $g_1(k)$ once we specify $c_1(t)$. Indeed, we can show [see Appendix~\ref{sec:appendixNonLinear} for details] that any \textit{physical} decay $c_1(t)$ can be realized  choosing the magnitude of the coupling to be
\begin{align}
    \abs{g_1(k)} =\sqrt{\frac{\vg}{\pi}\Im{\laplace{c}_1^{-1}(\Delta(k))}}\,,
    \label{eq:magnitudeCoupling}
\end{align}
where we introduced the Laplace transform $\laplace{c}_1$ of the decay as
\begin{equation}
    \laplace{c}_1(\Delta) := -i\int_0^{+\infty}\dd{t}{c}_1(t)e^{i\Delta t}\,.
\end{equation}
Eq.~\eqref{eq:magnitudeCoupling} specifies the meaning of {physical decay}, i.e., a decay $c_1(t)$ such that $\Im{\laplace{c}_1^{-1}(\Delta)}$ is non-negative.
If this condition is met, the assumed dynamics $c_1(t)$ can be created via passive interaction with a bath (the waveguide in the present case). The accessible dynamics are therefore also different from the ones created by time-dependent couplings~\cite{kurpiers_deterministic_2018, penas_improving_2023,pernas_shaping_2026}. 

The Laplace transform of the qubit dynamics can now be directly related to the wavevector components of the emitted pulse $\xi(k):=\lim_{t\rightarrow\infty}{c}_k(t)$ after full emission of the first qubit [cf. Eq.~\eqref{eq:eom} for a single qubit], 
\begin{equation}
    \xi(k)=g_1^*(k)\laplace{c}_1(\Delta(k))\,.
\end{equation}
Notice that $c_k(t)$ is defined in a rotating frame [see Eq.~\eqref{eq:wavefunction}] such that the corresponding pulse in real space, $\int\dd{k}e^{ikx}c_k(t)\xrightarrow{t\rightarrow\infty}\xi(x)$, is stationary after complete emission. Conversely, the emitted wavepacket in the stationary frame propagates through the waveguide while preserving its shape (due to the fact that the dispersion is linear).

We now wish to specify the phase $\arg{[g_1(k)]}$ of the coupling allowing for perfect state transfer.
For that it suffices to still consider the emission from the single qubit. Time-reversal symmetry dictates that if coefficients $c_1(t),\,c_k(t)$ solve the Schr\"odinger equation, so do $c^*_1(-t),\,c^*_k(-t)$ upon replacing $g_1(k)\rightarrow g_1^*(k)$ in the Hamiltonian. In other words, a pulse with Fourier amplitudes $\xi^*(k)$ is perfectly absorbed by a qubit with couplings $g_1^*(k)$. If we impose 
\begin{equation}
    \xi^*(k)=\xi(k)
\end{equation} (equivalent to $\xi(x)=\xi^*(-x)$ in real space) the pulse emitted by the qubit with coupling $g(x)$ can therefore in principle be absorbed by a subsequent qubit with coupling $g_1^*(k)$. 
For a given qubit decay the above condition is (up to a global phase) uniquely ensured by
\begin{equation}
    g_1(k):=|g_1(k)|e^{i\phi(k)}, ~\mathrm{with}~ \phi(k) = \arg\left[\laplace{c}_1(\Delta(k))\right]\,.
    \label{eq:phasePrescription}
\end{equation}
Now, if we place a second qubit at a distance $d$ with coupling points $x^{(2)}_i=d-x^{(1)}_i$ [see Fig.~\ref{fig:1}(a)] and coupling strengths $g^{(2)}_i=g^{(1)*}_i$, we can write $g_2(k) = e^{ikd}g_1^*(k)$. This means that the time-reversal-symmetric wavepacket will be reabsorbed by the second emitter after an additional propagation delay $d/\vg$. We stress that implementing Eq.~\eqref{eq:phasePrescription} for arbitrary qubit decays generated by a positive $|g(k)|$  will lead to the emission of time-reversal-symmetric pulses without changing the qubit dynamics.

The present conditions are derived under the assumption of a chiral waveguide. However, since we consider the decay timescale $\tau\gg\wq^{-1}$, the qubits couple only to wavevectors close to $\kq=\wq/\vg$. Therefore, while Eqs.~\eqref{eq:magnitudeCoupling} and \eqref{eq:phasePrescription} in principle uniquely determine $g(k)$, the coupling can still be suppressed around $k=-\kq$ with negligible influence on the dynamics. We will implement this explicitly for the following two examples.

\begin{figure}[tb]
    \centering
    \includegraphics[width=\linewidth]{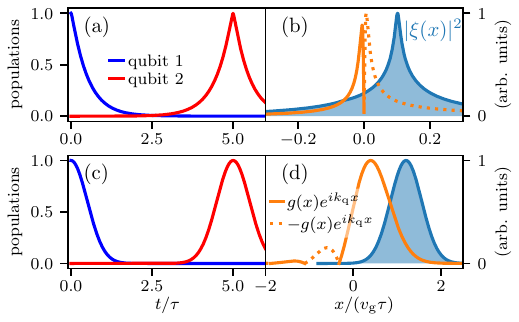}
    \caption{Quantum state transfer with wavevector-dependent couplings. (a) Populations of the qubits for an exponential decay 
    where we set $\gamma=2/\tau$. (b) Spatial coupling configuration (orange, solid lines for positive, dashed lines for negative values) which is real for the present cases after demodulating with $e^{i\kq x}$. This coupling profile leads to the emission of the associated time-reversal-symmetric single-photon pulse (blue, shifted to the right for better visibility). Panels (c,d) are analogous to panels (a,b) but for a Gaussian decay [see Eq.~\eqref{eq:Gaussian}]. \hll{All results are obtained for a bidirectional waveguide, i.e., the chiral emission is achieved solely through the giant atom design (see text for details).}} 
    \label{fig:fig2}
\end{figure}

\subsection{Examples: exponential and Gaussian decay}\label{sec:infinite:examples}
In the following we give two concrete examples for wavevector-dependent couplings that allow for perfect quantum state transfer. First, we consider an exponential decay described by
\begin{equation}
    {c}_1(t) = e^{- \gamma t/2}\,,~\mathrm{and~thus}~\laplace{c}_1(\Delta) = \frac{1}{\Delta+i\gamma/2}\,,\label{eq:exponential}
\end{equation}
that is typically associated to strictly Markovian dynamics. Such decay is achieved by a constant $\abs{g_1(k)}=\sqrt{\gamma/(2\pi)}$ according to Eq.~\eqref{eq:magnitudeCoupling} (we introduced $\gamma = 2/\tau$ as the inverse decay timescale to reproduce the expected results from the Markov approximation). Upon inserting Eq.~\eqref{eq:exponential} into Eq.~\eqref{eq:phasePrescription},
\hll{we arrive at the required wavevector-dependent coupling. We modify this momentum dependence to enforce chiral emission also when the giant atom is coupled to a bidirectional waveguide. To that end, we modify $g(k)$ as specified by Eq.~\eqref{eq:magnitudeCoupling} by multiplying with $\exp(-\varepsilon \vg|k-\kq|)$ for $\varepsilon\ll \tau$. This leaves $g(k)$ unchanged around $k=\kq$ and suppresses the coupling to $k=-\kq$ while allowing for good numerical convergence when calculating $g(x)$. Using this modified coupling, we find perfect transfer of the excitation 
in Fig.~\ref{fig:fig2}(a). }
In Fig.~\ref{fig:fig2}(b) we show the associated real-space coupling configurations and time-reversal-symmetric pulse that is emitted in this configuration, which then travels to the right. 
\hll{From the spatial configuration in Fig.~\ref{fig:fig2}(b) we can infer that the spatial extent of the giant atoms must be comparable to the width of the emitted pulse to achieve pulse shaping.}  \footnote{\hll{From Fig.~\ref{fig:fig2}(a) we see that after the second qubit absorbs the wavepacket, the latter is reemitted as a direct result of time-reversal symmetry. Since the emission timescale is identical to the absorption one, the state can in principle be stored by applying a $\pi$-pulse faster than this timescale in a $\Lambda$-type system~\cite{cirac_quantum_1997}.}}

As a second example, we consider a Gaussian decay, given by
\begin{align}
    c_1(t) = e^{-t^2/\tau^2}\,,    \label{eq:Gaussian}
\end{align}
and~thus
\begin{align} \laplace{c}_1(\Delta)=\tau\frac{\sqrt{\pi}}{2i}\exp\left(-\frac{\Delta^2\tau^2}{4}\right)\left[1+\mathrm{erf}\left(\frac{i\Delta\tau}{2}\right)\right]\,,
\end{align}
where $\mathrm{erf}(z)$ is the error function. For the Gaussian decay the coupling $g_1(k)$ is naturally restricted to narrow ranges around $\pm\kq$. A spatial configuration that chirally couples to the waveguide is then achieved analogously to the exponential case. The resulting dynamics is shown in Figs.~\ref{fig:fig2}(c) and (d) where we again find perfect transfer. Note that Eq.~\eqref{eq:phasePrescription} can be directly translated to perfect transfer in a closed loop waveguide~\cite{tian_static_2021} with two qubits opposite to one another. In that case, $|g_1(k)|$ has to be reduced by a factor of $\sqrt{2}$ in Eq.~\eqref{eq:magnitudeCoupling} to account for the second decay channel. 

\section{Finite number of legs}\label{sec:finiteLeg}
The considerations of the previous section have shown that in the limit of many legs perfect state transfer can be achieved with few constraints. In the following, we show that similar functionality can be achieved with a \textit{finite} number of coupling points. We note that as opposed to typical giant atom setups, in our case  these points need to be spaced by a distance on the order of the emitted pulse width $\sim\vg\tau$ to allow for shaping the pulse, i.e., the delays between the legs need to be non-negligible. We will justify that this is feasible in current experiments in Sec.~\ref{sec:implementations}. 

A finite leg implementation could naively be obtained by sampling equidistant points from the analytical configuration derived earlier (see Appendix~\ref{sec:appendixChirality} for an example). However, this procedure cannot be expected to yield the perfect transfer since we can only approximate the continuous distributions. In that context, it is also not clear which continuous coupling distribution (corresponding to a specific qubit decay) to discretize for the best transfer fidelity.

Therefore, instead of targeting a concrete qubit decay as in the previous section, we now numerically seek for the maximally achievable state transfer fidelity 
\begin{equation}
P_{2,\mathrm{max}} = \max_{\left\lbrace x^{(1)}_i, g^{(1)}_i, t\right\rbrace} |c_2(t)|^2
\label{eq:optimization}
\end{equation}
for a given number of coupling points, and optimize over leg positions, coupling strengths, and the time when the transfer is achieved. For the optimization we assume a chiral coupling to the waveguide and remove this requirement later on (see also Appendix~\ref{sec:appendixChirality}). In analogy to the case of infinitely many legs, we enforce time-reversal symmetry of the couplings of the two qubits, $g_2(k) = e^{ikd} g_1^*(k)$, and only optimize the first qubit's configuration, i.e., the second qubit's parameters are again derived as $x^{(2)}_i=d-x^{(1)}_i$ and $g^{(2)}_i=g^{(1)*}_i$. 
\begin{figure}[tb]
    \centering
    \includegraphics[]{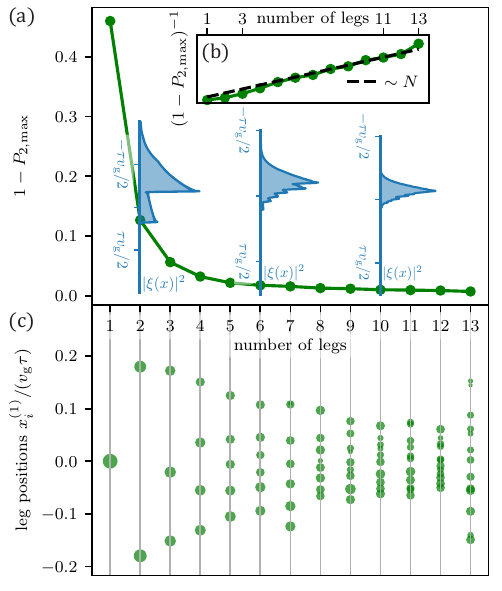}
    \caption{(a) State transfer infidelity as a function of the number of legs. We optimize the leg positions $x^{(1)}_i$ and coupling strengths $g^{(1)}_i$ under the constraint $\pi\sum_i|g^{(1)}_i|^2/\vg=\tau^{-1}$, where $\tau$ is the emission timescale, and ensuring that the qubits remain in a serial configuration. For $N=2, 6$ and 10 legs we plot the resulting single-photon pulse shape $|\xi(x)|^2$. (b) Further analysis reveals a scaling $1-P_{2,\mathrm{max}}\sim 1/ N$ with the number of legs $N$ (see linear fit).  (c) Leg positions associated to the optimization results in (a). The area of each point is associated to the coupling strength $|g^{(1)}_i|$.}
    \label{fig:fig3}
\end{figure}
Since there are no restrictions on the pulse shape, we have to introduce constraints to avoid pathologically large couplings. We do so by bounding the couplings by a fixed decay timescale $\tau$, $\pi\sum_i|g^{(1)}_i|^2/\vg=\tau^{-1}$, throughout the optimization. Specifically, this sets the initial decay of $|c_1(t)|$ of the giant atom before any partial pulse emitted from one leg reaches the next neighboring leg. Due to a scale invariance in the equations of motion our optimization results can be mapped to any value of the above constraint. \hll{In Appendix~\ref{sec:appendix:optimization}, we give details about the scale invariance and the optimization procedure.}

Fig.~\ref{fig:fig3}(a) shows the results of the optimization from 2 up to 13 legs of the giant atom, where we also added the analytical result for a single leg~\cite{stobinska_perfect_2009, wang_efficient_2011}. Already when adding a second leg, the attainable transfer fidelity is boosted from 54\,\% to around 87\,\%. With increasing number of legs, the transfer fidelity approaches unity, with the residual error of 1\,\% reached for around 10 legs.  The blue insets in Fig.~\ref{fig:fig3}(a) show the single photon pulses emitted from the giant atom setups for 2, 6 and 10 legs. The number of legs clearly reflects in the number of spikes of the single photon pulse owing to the discreteness of the couplings. 

With increasing number of legs the pulses become more time-reversal-symmetric which reflects in the transfer fidelity. In Fig.~\ref{fig:fig3}(b) we plot the inverse of $1-P_{2,\mathrm{max}}$ as a function of the number of legs $N$. This allows us to infer a scaling of the transfer error $\mathcal{O}(N^{-1})$. Perfect transfer, in turn, is only possible asymptotically for $N\rightarrow\infty$. This can be understood by considering that before the emissions from different legs interfere, each leg will start by independently emitting an exponentially shaped partial wavepacket, which does not comply with time-reversal symmetry. 
Fig.~\ref{fig:fig3}(c) shows the leg positions and coupling strengths obtained from our optimization in Fig.~\ref{fig:fig3}(a). They confirm that the spatial extent of the giant atoms must be on the order of magnitude associated with the decay rate $\sim \vg\tau$, which implies that the dynamics must be non-Markovian to favor state transfer. In fact, the clear separation between the wavelength scale and the leg spacing allows to make the giant atom emission inherently chiral by doubling the number of legs with appropriate phases (see Appendix~\ref{sec:appendixChirality}) in analogy to the two-leg case~\cite{ramos_non-markovian_2016,guimond_unidirectional_2020,wang_chiral_2022, joshi_resonance_2023, kannan_-demand_2023,almanakly_deterministic_2025,suarez-forero_chiral_2025}. \hll{We note that, while such coupling phases are often achieved by flux modulation of a Josephson junction~\cite{wang_chiral_2022}, Ref.~\cite{xu_chirality_2026} recently proposed a passive solution by including an individual delay line for each coupling point.} Further, when chiral coupling to the waveguide is achieved otherwise, the coupling strengths can be chosen real as their phases can be engineered by spatial shifts [see Eq.~\eqref{eq:realCouplingsPrescription}]. We also note that beyond optimizing for the state transfer, giant atoms may in principle also be optimized towards the emission of arbitrary pulse shapes (see Appendix~\ref{sec:appendixPulses}).

\section{Experimental considerations}\label{sec:IV}
\begin{figure*}
    \centering
    \includegraphics[]{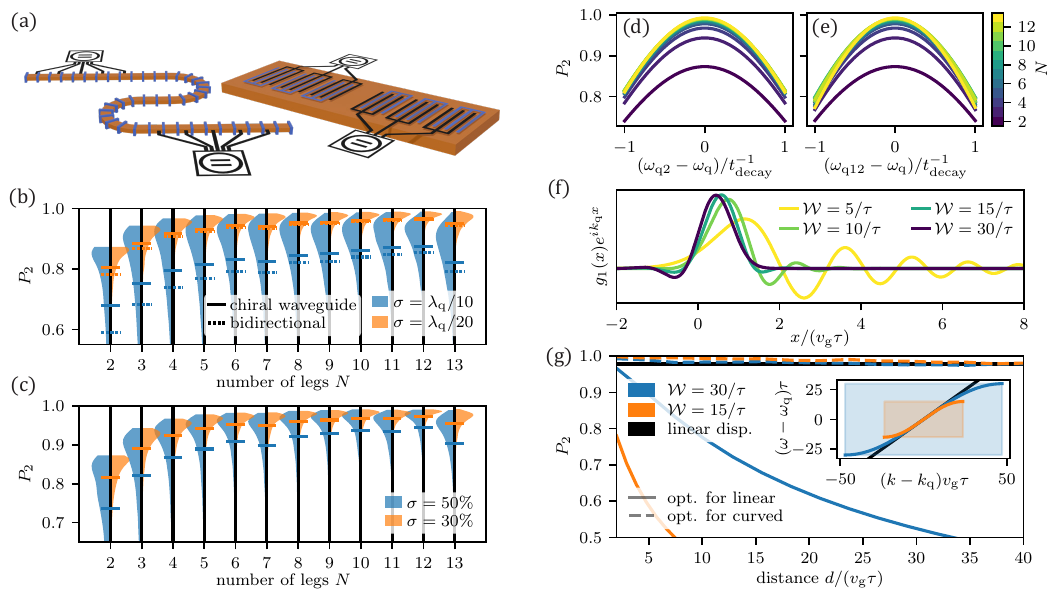}
    \caption{(a) Potential platforms to implement the transfer protocol using giant atoms coupled to a slow-light waveguide implemented in a cavity array (left) or surface acoustic waves (right).   
    (b) Sensitivity against random deviations of the leg positions with respect to the optimal configuration. The optimized leg positions of Fig.~\ref{fig:fig3} are displaced by Gaussian noise of width $\sigma=\lambda_\mathrm{q}/20$ (orange) and $\sigma=\lambda_\mathrm{q}/10$ (blue). The violin plot illustrates the resulting distributions for the transfer fidelity $P_2=\mathrm{max}_t|c_2(t)|^2$, where the \hll{horizontal} bar indicates the mean of the respective distribution. \hll{In analogy to Fig.~\ref{fig:fig3}, the results are obtained under the assumption of a chiral waveguide. We show the means for the implementation in a bidirectional waveguide with $2N$ legs according to Eq.~\eqref{eq:chiralPrescription} as dotted horizontal bars.} \hll{(c) Sensitivity to random deviations in the coupling strengths. The deviations are introduced analogously to panel (b), but for relative fluctuations of the optimal coupling strengths with Gaussian widths of $50\%$ or $30\%$. (d) Sensitivity to shifts in the frequency of a the receiving qubit $\omega_\mathrm{q2}$. For details of the frequency scale $t_\mathrm{decay}^{-1}$ see text. (e) Analogous to panel (d), but for correlated frequency shifts of both qubits with frequency $\omega_\mathrm{q12}$. (f)} Spatial coupling distributions for perfect chiral state transfer for a Gaussian decay and the curved dispersion of Eq.~\eqref{eq:sineDispersion}. The wavevector dependence is derived from Eqs.~\eqref{eq:curved-disp-gk} and \eqref{eq:curved-disp-phi} and the spatial distributions calculated by numerically inverting Eq.~\eqref{eq:spatialCouplings}. We consider the profiles for  various bandwidths $\mathcal{W}$.
    \hll{(g)} Giant atom state transfer in a curved dispersion. For the exemplary case of 5 legs, the state transfer fidelity as a function of qubit distance is studied in a curved dispersion of Eq.~\eqref{eq:sineDispersion} for two different bandwidths $\mathcal{W}$ (see inset). Using the leg setup from Fig.~\ref{fig:fig3} (optimized for a linear dispersion), the transfer fidelity quickly decreases (solid lines). Optimizing for each distance and dispersion bandwidth (dashed lines) the optimal transfer fidelity is recovered or even exceeded. }
    \label{fig:5}
\end{figure*}
In the following, we discuss how state transfer with giant atoms may be realized experimentally across different platforms. First, in Sec.~\ref{sec:implementations} we discuss potential experimental realizations of our approach. Then, we discuss the performance of our protocol in the presence of deviations from our model as they come with such realizations: In Sec.~\ref{subsec:legdisorder} we consider the sensitivity to disorder in the leg positions. Finally, in Sec.~\ref{subsec:nonlindisp} we study the effect of a nonlinear dispersion $\omega(k)$ of the waveguide.

\subsection{Experimental platforms}\label{sec:implementations}
 Our state transfer proposal applies to giant atoms with leg spacing on the order of the decay timescale $\sim \vg\tau$. While this is not the regime of operation of most giant atom implementations, we argue in the following that our approach can be implemented by combining slow propagation speeds with fast decay rates (i.e., strong enough coupling). Specifically, we consider the platforms of superconducting qubits coupled to surface acoustic waves (SAWs) and slow-light waveguides realized via coupled cavity arrays (CCAs).

SAWs represent the original platform in which the concept of giant atoms was introduced~\cite{gustafsson_propagating_2014}. 
The phonons travel at speeds around five orders of magnitude smaller than the speed of light~\cite{hadfield_quantum_2016}. For qubits with operating frequency in the $\si{\giga\hertz}$ range, the typical wavelengths are of $\sim1\si{\micro\meter}$~\cite{bienfait_phonon-mediated_2019}.
The qubit-SAW coupling can be strong enough to allow for decay times on the order of $\tau\sim 10 \si{\nano\second}$~\cite{gustafsson_propagating_2014, bienfait_phonon-mediated_2019}, corresponding to pulse lengths of $40\si{\micro\meter}$, i.e., giant atom legs would need to be spaced on this length scale. This leaves a sizeable scale difference to the wavelength, while still allowing the realization of giant atoms within a chip-scale device. \hll{Reference~\cite{bienfait_phonon-mediated_2019} implements state transfer by active modulation of qubits in an acoustic Fabry-Pérot resonator with \SI{2}{\milli\meter} mirror spacing. Such a system could serve as a proof-of-principle implementation of our proposal.}

The propagation speed may also be reduced by two to three orders of magnitude in microwave slow-light waveguides~\cite{mirhosseini_superconducting_2018}. The corresponding wavelengths hence are $\sim100\si{\micro\meter}$.
Such waveguides have been shown to reach qubit emission timescales as low as $\tau\sim 7.5\si{\nano\second}$ (at a group index of around $650$), corresponding to wavepacket sizes of few millimeters compatible with on-chip design~\cite{ferreira_collapse_2021}. Alternatively, high kinetic-inductance metamaterials have been shown to allow for the realization of compact CCAs~\cite{jouanny_high_2025}. This enables to couple giant atoms to $\sim\!10$ cavities at the same time \cite{jouanny_superstrong_2025}, while being able to control the strength of the coupling to each cavity. \hll{In Appendix~\ref{sec:appendix:cca} we discuss the implementation of our protocol with CCAs, and propose and characterize explicitly setups with four and five coupling points based on recent experiments~\cite{jouanny_high_2025,jouanny_superstrong_2025}}.

\subsection{\hll{Robustness} against random deviations in the leg positions}
\label{subsec:legdisorder}
The success of the state transfer is particularly tied to the precise positioning of the coupling points, as this determines the interference leading to a time-reversal-symmetric pulse. This positioning thus needs to be precise within the length scale of the wavelength of the qubit transition. 
For this reason, in Fig.~\ref{fig:5}(b) we show the influence of deviations from the optimized positions of the legs on the state transfer performance~\cite{han_giant_2026}. Starting from the optimization results presented in Fig.~\ref{fig:fig3}, we introduce disorder by randomly shifting each leg by a distance $\delta x$, drawn from a zero-mean Gaussian distribution of width $\sigma$, and simulate the resulting dynamics \hll{under the assumption of a chiral waveguide}. 
The transfer fidelity distributions after sampling over many of these noisy implementations are shown in Fig.~\ref{fig:5}(b). While generally good transfer fidelities retain the highest probabilities in the distributions, we conclude that a precision of $\lambda_\mathrm{q}/20$ in the positions of the legs is required for the mean fidelity to stay within $\approx 5\%$ of the maximum achievable fidelity  for a given number of legs. \hll{Note that for SAWs, which are the smallest structures we discuss in Sec.~\ref{sec:implementations}, $\lambda_\mathrm{q}/20$ corresponds to a placement precision of \SI{50}{\nano\meter}, as may be routinely achieved by electron-beam lithography~\cite{vieu_electron_2000}.} \hll{To also quantify the sensitivity when coupling to a bidirectional waveguide and enforcing chirality by doubling the number of legs, we repeat the analysis without the assumption of a chiral waveguide. To that end, we again start from the optimized results in Fig.~\ref{fig:fig3}, double the number of legs according to Eq.~\eqref{eq:chiralPrescription} and simulate the dynamics to get the mean transfer fidelity. We plot the means as dotted bars in Fig.~\ref{fig:5}(b), where the actual number of legs is now $2N$, i.e., the x-axis only indicates half of the number of physical legs. We find that for small fluctuations, doubling the number of legs for chirality introduces only negligible additional sensitivity to fluctuations in the leg positions.}

\hll{
\subsection{\hll{Robustness} against random deviations in the coupling strengths}
\label{subsec:couplingdisorder}
Besides deviations in the leg positions, also the individual legs' coupling strengths may be subject to deviations in the manufacturing process. To also study the influence of such deviations, we proceed analogously to Sec.~\ref{subsec:legdisorder}. Starting from the optimal configurations of Fig.~\ref{fig:fig3}, we draw relative fluctuations for each leg from a Gaussian distribution of width $\sigma$ and simulate the resulting transfer for many different random configurations. We plot the resulting distributions and their mean in Fig.~\ref{fig:5}(c). Here, we find that the transfer fidelity stays within about $5\%$ of its optimal value when the width of the relative variations does not exceed $30\%$. This scale of deviation can be achieved in current experiments (see for example Ref.~\cite{bienfait_phonon-mediated_2019}, where two qubits are manufactured with relative deviation below 17\% in the maximal coupling).}

\hll{\subsection{\hll{Robustness} against deviations in the qubit frequencies}
\label{subsec:frequencydisorder}
From Figs.~\ref{fig:fig2}(b) and \ref{fig:fig2}(d), we can see that the phases of the spatial coupling configuration depend on the qubit frequency $\wq$ via $\kq$. Therefore, we should expect that also the discrete configurations we optimized for in Fig.~\ref{fig:fig3} are sensitive to deviations in the qubit frequency. Before studying this sensitivity, we need to consider the relevant scale for frequency fluctuations for our setups. We recall that, while we set a constraint based on the timescale $\tau$, this constraint did not precisely fix the actual decay timescale, which is found to vary for different numbers of legs in Fig.~\ref{fig:fig3}. At the same time, the decay timescale sets the linewidth of the transition, and therefore the relevant frequency scale. To find the actual decay timescale, we therefore simulate the decay for each $N$ in Fig.~\ref{fig:fig3}, and determine the time $t_\mathrm{decay}$ when the excitation probability of the qubit has dropped to $|c_1(t_\mathrm{decay})|^2=1/e^2$. This definition is motivated by the fact that for an exponential decay $c_1(t)=e^{-t/\tau}$ we would precisely end up with the linewidth $\tau^{-1}$. Taking $t_\mathrm{decay}^{-1}$ as the frequency scale, we now study two different scenarios: First, in Fig.~\ref{fig:5}(d) we consider a frequency shift of the second qubit's frequency $\omega_\mathrm{q2}$ only. Second, in Fig.~\ref{fig:5}(e) we study the sensitivity when both qubits share the same frequency $\omega_\mathrm{q12}$, which is different from the frequency $\wq$ assumed at design time. In both cases, we find that the first relevant contribution to the frequency dependence is $\mathcal{O}([\omega-\wq]^2/t_\mathrm{decay}^{-2})$, i.e., the protocol is robust to small deviations in frequency. For larger deviations, the fidelity stays above 95\% as long as the deviations are below $0.25t_\mathrm{decay}^{-1}$ and the number of legs $N\geq 4$. For qubits with $t_\mathrm{decay}\sim\SI{10}{\nano\second}$~\cite{bienfait_phonon-mediated_2019,ferreira_collapse_2021} this would correspond to a stabilization of the qubit frequency to \SI{25}{\mega\hertz}, which is straightforwardly achieved \cite{vepsalainen_improving_2022,  zhang_high-performance_2022}.}

\subsection{Sensitivity to and correction of nonlinear dispersion effects}
\label{subsec:nonlindisp}
In many systems, the dispersion law $\omega(k)$ is not a linear function of the photon wavenumber $k$. This entails that photons at different frequencies propagate at different speeds, given by $v(k)=\partial\omega/\partial k$. 
Thus, the shape of the pulse emitted by a qubit is distorted during its propagation, which raises the question whether we may extend our approach of Sec.~\ref{sec:finiteLeg} [see Eq.~\eqref{eq:phasePrescription}] to regain {perfect state transfer} for a generic nonlinear dispersion. In the following, we show that this is indeed possible. Differently from Sec.~\ref{sec:infiniteLegs},  we now allow $\omega(k)$ to be a nonlinear, continuous band of chiral photons (i.e., $v(k)>0$). In this situation, the conditions for the modulus and phase of the coupling allowing for perfect state transfer can be generalized to (see Appendix~\ref{sec:appendixNonLinear})
\begin{align}
    \label{eq:curved-disp-gk}
    \abs{g_1(k)} &= \sqrt{\frac{1}{\pi\rho(k)} \Im{\laplace{c}_1^{-1}(\Delta(k))}}\,,\\
    \phi(k) &= \arg{\qty[\laplace{c}_1(\Delta(k))]} + \qty[kd-\omega(k)T]/2\,.\label{eq:curved-disp-phi}
\end{align}
Here, $\rho(k) = \abs{\partial k/\partial\omega}$ is the density of states and $T$ is the \textit{transfer time}, i.e., the instant at which the population will be completely transferred to the second qubit. The phase correction Eq.~\eqref{eq:curved-disp-phi} can be understood as the phase needed to tailor a specific chirped pulse, that compensates for the distortion during propagation~\cite{casulleras_remote_2021}. While the time $T$ is a free parameter, it is bounded by $T\geq d/v_{\rm max}$, where $v_{\rm max}=\max_k |\partial\omega/\partial k|$ is the speed of the fastest-moving photons. Note that the phase profile in Eq.~\eqref{eq:curved-disp-phi} now has an explicit dependence on distance. Also notice that $\omega(k) = \vg |k|$ together with the chiral coupling assumption leads to $\rho(k) = \vg^{-1}$, i.e., we recover the coupling to a linear waveguide setting with $T=d/\vg$. To showcase the influence of the curved dispersion on the spatial couplings necessary for perfect transfer, we consider a Gaussian decay [see Eq.~\eqref{eq:Gaussian}] in a sinusoidal dispersion (similar to the dispersion in a CCA)
\begin{equation}
    \omega(k)=\wq+\mathcal{W}\sin([\vg |k|-\wq]/\mathcal{W}),
    \label{eq:sineDispersion}
\end{equation}
with bandwidth $\mathcal{W}$ centered at the qubit frequency $\wq$. In Fig.~\ref{fig:5}\hll{(f)}, we plot the spatial couplings for different bandwidths $\mathcal{W}$ for a distance of $d=30\vg\tau$. For large bandwidths, the dispersion behaves similar to a linear one, i.e., we recover the spatial profile in Fig.~\ref{fig:fig2}(d). For smaller bandwidths, the coupling configuration is found to be more extended in space. This is necessary to account for the more pronounced wavevector dependence according to Eq.~\eqref{eq:curved-disp-phi}. 

The exact results in Eq.~\eqref{eq:curved-disp-gk} can be implemented with a giant atom in the limit of an infinite number of legs. For a finite number of legs, however, we can proceed analogous to Sec.~\ref{sec:finiteLeg} and run an optimization to find the optimal coupling configuration. When using the parameters of Fig.~\ref{fig:fig3}, which are optimized for a linear dispersion, for state transfer in the curved dispersion Eq.~\eqref{eq:sineDispersion} the transfer fidelity rapidly deteriorates with increasing distance between the qubits [see Fig.~\ref{fig:5}(d)]. The stronger the nonlinearity, the more pronounced we find the deterioration to be, since pulse distortion effects become stronger. Starting from the optimal configuration found for the linear waveguide, we now repeat the optimization of the leg positions and couplings strengths [cf.~Eq.~\eqref{eq:optimization}] for the nonlinear dispersion of Eq.~\eqref{eq:sineDispersion}. We find that the transfer performance achieved in the linear case can not only be restored in accordance with the analytical results, but even exceeded, as the dispersion may aid in the state transfer~\cite{kuang_perfect_2025}.

\section{Conclusions}\label{sec:conclusion}
We have shown that giant atoms provide a natural platform for realizing deterministic passive quantum state transfer. Specifically, we have demonstrated that it is possible to tailor the emitters' nonlocal coupling to an electromagnetic environment such that the wavepacket emitted by spontaneous decay is time-reversal-symmetric, a property that enables perfect state transfer and is typically realized by active control. In the continuum limit, we established general analytical conditions linking arbitrary qubit decays to wavevector-dependent couplings that guarantee perfect reabsorption. For realistic implementations with a finite number of coupling points, we found that high fidelities can be achieved with only a modest number of legs, with performance improving systematically toward unity as the number of coupling points increases. The protocol remains robust against moderate imperfections in the leg positions and can be applied also to state transfer in nonlinear dispersions as they result from structured photonic environments. 

Our results highlight a conceptual shift in quantum state transfer: the usually assumed temporal profile of the qubit-waveguide coupling can be directly encoded into a fixed spatial pattern of coupling points. \hll{This demonstrates that complex dynamical control can be included into hardware design, thus offering a novel design paradigm for quantum technology applications. We expect that the approach will be beneficial in practice also when combined with established active state transfer schemes, thus reducing requirements on the giant atoms design as well as on the pulse shaping.}
\hll{Our theoretical framework and designs may further be useful also beyond giant atom implementations, in particular to any platform where one can engineer non-trivial frequency-dependent coupling, e.g., coupled-cavity based nodes~\cite{tian_static_2021}.}
Extending the present framework to multi-qubit and multi-photon settings could enable the transfer of multipartite entanglement using tailored giant-atom geometries. Further, the ability to engineer emission profiles suggests applications beyond state transfer, such as passive pulse shaping, quantum memories, or the realization of effective non-Markovian reservoirs with programmable properties.

\begin{acknowledgments}
\hll{We thank Pasquale Scarlino for helpful discussions.}
FC and EDB acknowledge financial support from European Union – NextGenerationEU through projects: Eurostart 2022
‘Topological atom-photon interactions for quantum technologies’; PRIN 2022–PNRR No. P202253RLY
‘Harnessing topological phases for quantum technologies’; THENCE–Partenariato Esteso
NQSTI–PE00000023–Spoke 2 ‘Taming and harnessing decoherence in complex networks’.
This research was funded in part by the Austrian Science Fund (FWF) [10.55776/COE1],     [10.55776/PAT1177623], [10.55776/PIN3404324] and the European Union – NextGenerationEU. For open access purposes, the author has applied a CC BY public copyright license to any author accepted manuscript version arising from this submission
This research is part of the Munich Quantum Valley, which is supported by the Bavarian state government with funds from the Hightech Agenda Bayern Plus.
AGT acknowledges support from the CSIC Research Platform on Quantum Technologies PTI-001, Spanish project Proyecto PID2024-162384NB-I00 financiado por MICIU/AEI/10.13039/501100011033 y por FEDER,UE, from the
QUANTERA project MOLAR with reference PCI2024153449 and funded MICIU/AEI/10.13039/501100011033 and by the European Union, the Programa Fundamentos FBBVA through the grant EIC24-1-17304.
The computational results have been achieved using the Austrian Scientific Computing (ASC) infrastructure. 
\end{acknowledgments}

\begin{appendix}
\section{Chiral Emission \& Coupling Phases}\label{sec:appendixChirality}
\begin{figure*}
    \centering
    \includegraphics[]{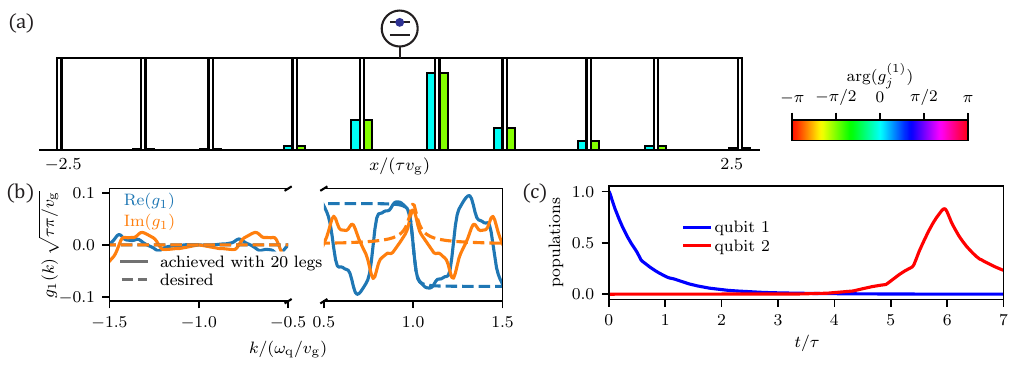}
    \caption{Directional emission for quantum state transfer by doubling the number of legs. (a) Setup of a giant atom coupled at equidistant points in a window of 5$\,\vg\tau$ with coupling strengths sampled from Eq.~\eqref{eq:spatialExpCoupling}. After sampling 10 points, they can be made real-valued by shifts on the wavelength scales [see Eq.~\eqref{eq:realCouplingsPrescription}]. Directionality with only two different phases is then achieved by doubling the number of points according to Eq.~\eqref{eq:chiralPrescription}. The height of the bars indicates coupling strength and the color indicates the coupling phase. Bars are shifted to the left or right of their respective coupling point for better visibility. (b) Wavevector-dependent coupling resulting from the configuration in (a). While the desired coupling is achieved locally for positive $k$ modes, the negative $k$ modes couple with negligible coupling strength on resonance. (c) Qubit dynamics resulting for the configuration in (a). For the plot we use $\wq\tau = 50$ such that leg spacing on the wavelength scale is still visible on the decay rate scale.}
    \label{fig:chiral}
\end{figure*}
In Sec.~\ref{sec:infiniteLegs} we showed that for a continuous spatial coupling distribution we can intrinsically achieve a chiral coupling. Here, we give a simple recipe how to also get this feature for a finite number of coupling points. It has been shown previously that chiral emission can be achieved for one leg when adding a second leg at distance $\lambda/4$ with an additional phase of $\pi/2$~\cite{ramos_non-markovian_2016,guimond_unidirectional_2020,wang_chiral_2022, joshi_resonance_2023, suarez-forero_chiral_2025}, as can be engineered, e.g., in superconducting qubits~\cite{joshi_resonance_2023}. The analogous approach is applicable to our setup when adding a second leg for each existing one,
\begin{equation}
    g_\mathrm{1,chiral}(k) = \sum_jg^{(1)}_j\left(e^{ikx^{(1)}_j}+ie^{ik(x^{(1)}_j+\lambda_\mathrm{q}/4)}\right)/2\,,
    \label{eq:chiralPrescription}
\end{equation}
which enforces the conditions $g_\mathrm{1,chiral}(k_\mathrm{q}) =g_1(k_\mathrm{q})$ and $g_\mathrm{1,chiral}(-k_\mathrm{q}) =0$. While this approach in principle modifies the precise wavevector dependence of the couplings, these modifications happen on a scale $\Delta k \sim \wq/\vg$. In contrast, the relevant  scale for the creation of time-reversal-symmetric pulses is given by $\Delta k\sim 1/(\vg\tau)$, which is much smaller than the one inducing chirality. Therefore, the design of couplings can in principle be performed under the assumption of chiral coupling and translated to a bidirectional waveguide later on. 

We demonstrate the applicability of the scheme by first analytically deriving  a continuous spatial coupling distribution for an exponential decay, then sampling this continuous function to arrive at discrete couplings and finally doubling the number of legs to arrive at chiral emission. First, we derive the real space coupling distribution for an exponential decay [see Eq.~\eqref{eq:exponential}] from Eq.~\eqref{eq:phasePrescription} under the assumption of a chiral waveguide, i.e., \textit{without} the regularization $\exp(-\varepsilon \vg|k-\kq|)$ employed for Fig.~\ref{fig:fig2}(b),  
\begin{align}
    g_1(x)
    &=\frac{e^{-i\kq x}}{(\tau\pi)^{3/2}\sqrt{\vg}}
\left[K_{0}\!\left(\frac{|x|}{\vg\tau}\right)+\operatorname{sgn}(x)\,K_{1}\!\left(\frac{|x|}{\vg\tau}\right)\right]\,,
\label{eq:spatialExpCoupling}
\end{align}
where $K_\alpha$ are Bessel functions of the second kind of order $\alpha$ and we parametrize the decay rate by $\tau=2/\gamma$ for notational consistency. While this coupling may mediate perfect transfer in a chiral waveguide with only right-propagating photons, it will not work in a bidirectional waveguide since without regularization, $|g_1(k)|$ remains constant at $k=-\kq$  such that the qubit also emits into left-propagating modes.  
From the analytical form, we can confirm that variations of the spatial coupling profile indeed happen on a scale $\sim \vg\tau$.
To represent this continuous coupling configuration by a finite number of legs, we sample 10 equidistant points $x_i^{(1)}$ in a $5\vg\tau$ window and local couplings $g_i^{(1)}=\delta x\,g_1(x_i^{(1)})$ where $\delta x$ is the spacing of neighboring points. In general, the sampled points will be complex-valued. However, due to the above mentioned separation of wavevector scales, any set of complex-valued discrete couplings may be translated to a real-valued one with negligible influence on the dynamics by replacing
\begin{equation}
    x_j^{(\alpha)}\rightarrow x_j^{(\alpha)}+\arg(g_j^{(\alpha)})/\kq\,,\quad g_j^{(\alpha)}\rightarrow |g_j^{(\alpha)}|\,.
    \label{eq:realCouplingsPrescription}
\end{equation}
In particular, this applies to systems where chirality is achieved otherwise, e.g., in quantum Hall analogues~\cite{de_bernardis_chiral_2023}.
We can now apply the prescription of Eq.~\eqref{eq:chiralPrescription} to double the number of coupling points to arrive at the configuration of Fig.~\ref{fig:chiral}(a), which due to Eq.~\eqref{eq:realCouplingsPrescription} only has two different coupling phases, $\phi=0$ and $\phi=\pi/2$ among all legs. In Fig.~\ref{fig:chiral}(b) we see that for positive momenta, the coupling distribution follows the desired one when being close to resonance. Oppositely, for negative momenta, the coupling is almost completely suppressed. In Fig.~\ref{fig:chiral}(c) we see that, while the dynamics is reminiscent of the scheme in Fig.~\ref{fig:fig2}(a) in the main text, the fidelity lags behind due to the finite number of legs. Furthermore, we do not reach fidelities comparable to Fig.~\ref{fig:fig3} as we use a non-optimized configuration.
\section{General theory of perfect state transfer in a photonic bath}
\label{sec:appendixNonLinear}
We consider two qubits coupled to a 1D photonic bath whose normal modes are created by operator $a_k^\dag$ and have a frequency $\omega(k)$. We assume that the dispersion law $\omega(k)$ forms a single continuous band in the interval $\comm{\omega_{\rm min}}{\omega_{\rm max}}$ parametrized by the continuous parameter $k$, i.e., there is a (potentially finite) bandwidth $\mathcal{W} = \omega_{\rm max}-\omega_{\rm min}$, and assume that $\omega_{\rm max}\geq\wq\geq\omega_{\rm min}$. Also, we assume that the bath is \textit{chiral}, i.e., it allows for photons propagating in a single direction. As discussed in Sec.~\ref{sec:infiniteLegs} and Appendix~\ref{sec:appendixChirality} this assumption can be relaxed later on. 
The Hamiltonian [see Eq.~\eqref{eq:Hamiltonian}] is excitation preserving and the dynamics thus restricted to the single excitation subspace [see Eq.~\eqref{eq:wavefunction}].
Our goal is to find conditions for which the first qubit, initially excited by a single excitation, decays completely by emitting a wavepacket which, after propagation in the bath, gets perfectly reabsorbed by the second qubit. Since the processes of emission and reabsorption will be separated in time, we can neglect for a moment the presence of the second qubit which allows us to study in a simpler way how to engineer a specific physical decay of the first qubit.
Using the resolvent operator approach~\cite{cohen-tannoudji_atom-photon_2008, gonzalez-tudela_markovian_2017}, the evolution of the qubit wave amplitude $c_1(t)$ can be expressed as
\begin{equation}
    \label{eq:resolvent-c1}
    c_1(t) = \frac{i}{2\pi} \int_{-\infty}^{+\infty} \dd \Delta \frac{e^{-i\Delta t}}{\Delta-\Sigma(\Delta+\wq+i0^+)}\,,
\end{equation}
where we introduced the self-energy $\Sigma(E)$ given by
\begin{equation}
    \label{eq:resolvent-self}
    \Sigma(E) = \int \dd k \frac{\abs{g_1(k)}^2}{E-\omega(k)} = \int \dd \omega\, \rho(\omega) \frac{\abs{g_1(k(\omega))}^2}{E-\omega}\,,
\end{equation}
and $\rho(\omega) = \abs{\partial k/\partial \omega}$ is the density of states.
We now define the Laplace transform of $c_1(t)$ as
\begin{equation}
    \label{eq:resolvent-laplace}
    \laplace{c}_1(\Delta) = -i\int_0^{+\infty} \dd t \, e^{i\Delta t} c_1(t)\,.
\end{equation}
Upon substituting Eq.~\eqref{eq:resolvent-c1} into Eq.~\eqref{eq:resolvent-laplace}, we are able to express the self energy as $\Sigma(\Delta+\wq+i0^+) = \Delta-\laplace{c}^{-1}_1(\Delta)$, whose imaginary part reads
\begin{equation}
    \Im{\frac{1}{\laplace{c}_1(\Delta)}} = -\Im{\Sigma(\Delta+\wq+i0^+)}\,.
\end{equation}
From Eq. \eqref{eq:resolvent-self}, we can express the imaginary part of the self-energy as
\begin{widetext}
     \begin{equation}
        \Im{\Sigma(E+i0^+)} = -\pi \rho(E) \abs{g_1(k(E))}^2\times
        \begin{cases}
            1\,, & \text{if } \omega_{\rm min}< E<\omega_{\rm max}\,,\\
            \frac{1}{2}\,, & \text{if } E = \omega_{\rm min}, E=\omega_{\rm max}\,,\\
            0\,, & \text{otherwise}\,.
        \end{cases}
    \end{equation}
With this in hand, we can compute the modulus of the coupling as
\begin{equation}
    \label{eq:resolvent-ge}
    \abs{g_1(k)} = \sqrt{\frac{1}{\pi\rho(\wq+\Delta(k))} \Im{\laplace{c}^{-1}_1(\Delta(k))}}\times
    \begin{cases}
        1\,, & \text{if } k(\omega_{\rm min})< k <k(\omega_{\rm max})\,,\\
        \sqrt{2}\,, & \text{if } k = k(\omega_{\rm min}), k(\omega_{\rm max})\,,
    \end{cases}
\end{equation}
\end{widetext}
where we introduced $\Delta(k)=\omega(k)-\wq$. Eq.~\eqref{eq:resolvent-ge} gives the shape of the coupling in $k$-space needed to engineer any \textit{physical decay} in a 1D chiral photonic bath. 

Now that we understand how to tailor the qubit emission by specifying the modulus of the coupling strength, we can reintroduce the second qubit and engineer perfect quantum state transfer by choosing the correct phase of the coupling. Notice that at long times, i.e., after the first qubit has completely emitted its excitation, the emitted wavepacket has an envelope given by
\begin{equation}
    \xi(k) = \lim_{t\rightarrow\infty} c_k(t) = g_1^*(k) \laplace{c}_1(\Delta(k))\,.
\end{equation}
Thus, at time $t$ the emitted field reads $\psi(k,t) = \xi(k) e^{-i\omega(k) t}$ and drives the second qubit as $g_2(k) \psi(k,t)$. Now, if we choose the second qubit's coupling to be $g_2(k) = e^{ikd} g_1^*(k)$ (corresponding to the time-reversed coupling shifted in space), the field $\psi(x,t)$ would be perfectly reabsorbed at time $T$ if it was driving the second qubit with a pulse $g_1^*(k)\xi^*(k)$. This means that, in order to have perfect reabsorption we have to impose
\begin{equation}
    g^*_1(k) \xi^*(k) = g_2(k) \xi(k) e^{-i\omega(k) T}\,.
\end{equation}
After some algebra, we end up with
\begin{equation}
    \arg\qty{g_1(k)} = \arg\{\laplace{c}_1(\Delta(k))\} + \frac{1}{2}\qty[kd-\omega(k) T]\,,
\end{equation}
which specifies the phase of the coupling profile in $k$-space.
Notice that the phase consists on two terms, the first coming from the physical decay of the qubit and the second from the nonlinear dispersion law. The latter compensates the distortion of the pulse envelope during its propagation. In other words, in a bath with a nonlinear dispersion law, perfect state transfer is achieved emitting a \textit{chirped pulse}, whose correct shape is then restored during the propagation~\cite{casulleras_remote_2021}. Indeed, if the dispersion is linear as $\omega(k) = \vg k$ and the transfer time is chosen to be $T=d/\vg$, this second contribution cancels out.

Once we know the full expression of the coupling, we can go back to real space using a continuous or discrete Fourier transform, depending on whether our bath is continuous (e.g., a waveguide) or discrete (e.g., a coupled cavity array) in the position coordinate.
We now study as an example the case of qubits coupled to a Coupled-Cavity Array (CCA). This will give us the chance to show how the analytical results derived above can be extended to a non-chiral bath.

\subsection{Coupled-Cavity Array}
\label{sec:appendix:cca}
A CCA is a system made by a set of equally-spaced cavities $(n=1,...,N)$, with spacing $a$, whose resonant frequency is set to $\omega_\mathrm{r}$. In its simplest realization, each cavity is coupled with its nearest-neighbor with a hopping rate $J<0$. In the thermodynamic limit, i.e., when the number of cavities $N\to\infty$, the dispersion law reads $\omega(k) = \omega_\mathrm{r}-2\abs{J}\cos{ka}$, where $\abs{k}\leq \pi/a$. Clearly, such a system is not chiral, since the group velocity $v(k) = 2\abs{J}a\sin{ka}$ is positive (negative) for $k>0$ ($k<0$). Nonetheless, we can still apply the analytical result derived earlier in the following way.

First of all, we assume that the first qubit couples only to modes having $k>0$ (right-moving photons). We then compute the coupling $g_1(k)$ following the recipe given above. We then modify such coupling configuration as
\begin{equation}
    \label{eq:CCA-filter}
    G_1(k) = g_1(k) \frac{1-e^{i(k+\kq)a}}{1-e^{2i\kq a}}\,,
\end{equation}
where $\kq  =\arccos{\qty(\wq/2J)}/a$ is the wavevector resonant with the qubit. Going back to real space, this entails
\begin{equation}
    G^{(1)}_n = \frac{g^{(1)}_n-e^{i\kq  a }g^{(1)}_{n+1}}{1-e^{2i\kq  a}}\,,
\end{equation}
which ensures chiral emission of the first qubit and perfect reabsorption by the second one.

This works because the relevant contribution to the emission process comes from momenta around the resonant momenta $\pm \kq $, on a scale of the order of $\Delta k \sim 1/(v(\kq )\tau a)$, being $\tau$ the decay time scale.
The expression in Eq. \eqref{eq:CCA-filter} filters out the coupling to the $-\kq $, responsible for emission in the wrong direction. In principle, this modifies the coupling in $k$-space on a scale given by $\kq $. Thus, as long as $\Delta k \ll \kq $, these modifications will not affect significantly the emission process. This entails that the decay time scale needs to be $\tau \gg 1/(v(\kq ) \kq  a)$. Notice that, since $\Im{\laplace{c}_1^{-1}(\Delta(k))}\sim 1/\tau$, this is nothing but a requirement of \textit{weak-coupling} of the qubits to the CCA, which stems from the fact \hll{that} the bandwidth is finite and so $\kq$ cannot be arbitrarily high.

\begin{figure*}[t]
    \centering
    \includegraphics[]{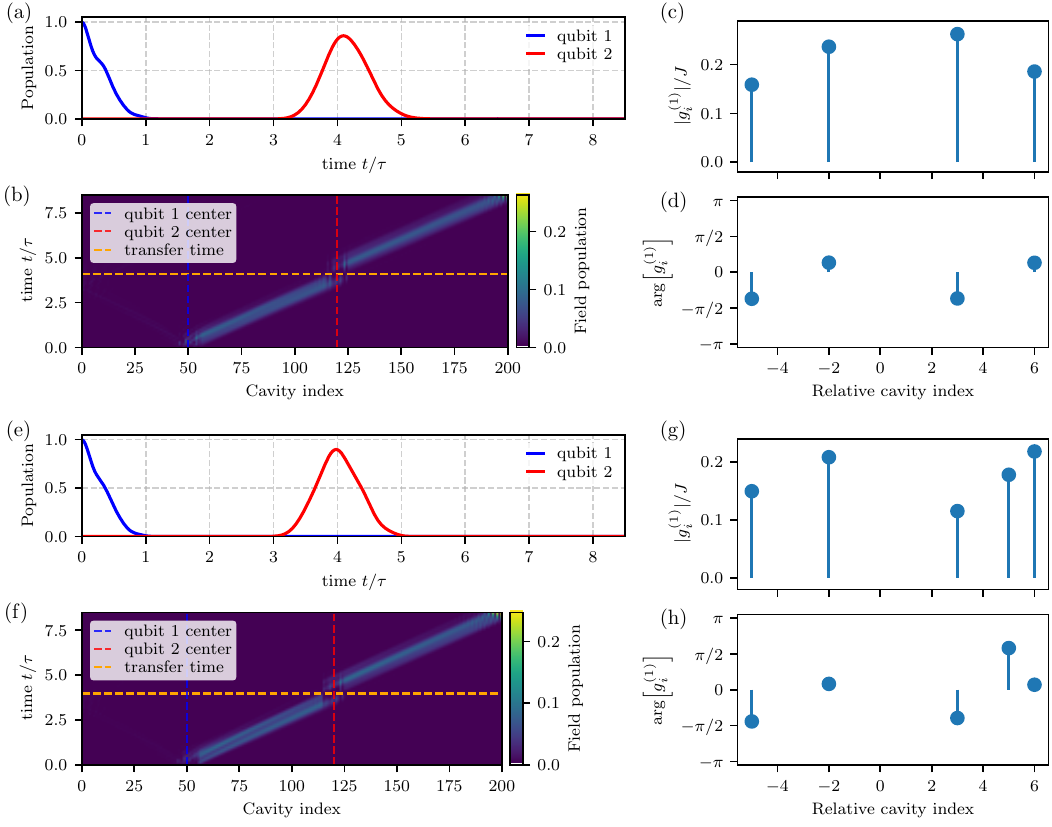}
    \caption{\hll{State transfer dynamics in a CCA for a giant atom with four (a-d) and five legs (e-h). (a,e) Qubit dynamics. The blue (red) curves represent the time evolution of the population of the first (second) qubit. (b,f) Populations in the cavities as a function of time. The orange dashed line corresponds to the instant in which the maximum population is transferred to the second qubit, while the blue (red) dashed line correspond to the center of the first (second) qubit. (c-d,g-h) Magnitudes (in units of $J$) and phases of the discrete coupling points.}}
    \label{fig:CCA}
\end{figure*}

\hll{Alternatively to employing the above prescription for chirality, we can also directly optimize for the transfer fidelity in a (bidirectional) CCA for a given number of coupling points. Below, we perform such an optimization for a realistic experimental implementation using transmon qubits coupled to high-impedance CCA~\cite{jouanny_high_2025}. In such a CCA, strong coupling of a giant atomic emitter to $\sim7$ cavities of the array has been recently demonstrated \cite{jouanny_superstrong_2025}. 
In this architecture, the physical body of the transmon can be shaped to engineer both the spatial positions and the magnitudes of the couplings to specific, distant cavities.
Furthermore, this structural design is fully compatible with parametric coupling schemes \cite{joshi_resonance_2023, yang_2026_many} enabling the imprinting of a specific static phase at each individual coupling point. 
By combining transmon shape engineering with parametric modulation, realizing a device where each giant atom couples to five or more distinct cavities with engineered complex phases is within reach.}

\hll{In Fig.~\ref{fig:CCA}, we present the simulated state-transfer dynamics between two qubits coupled to a common CCA via four and five coupling points. For these simulations, the qubits are resonant with the bare cavity frequency ($\omega_\mathrm{q}=\omega_\mathrm{r}$), so that they spectrally lie in the center of the passband, and the inter-cavity hopping rate is set to $J/2\pi = 250\,\si{\mega\hertz}$. Notice that we assume a positive hopping rate here to align with current experimental implementations.
Furthermore, we assume the CCA to comprise $N=200$ cavities and the emitters to be separated by a distance of $d=70$ cavities.
The coupling configurations used here were derived using an optimization approach analogous to the one employed for the linear waveguide [see Appendix \ref{sec:appendix:optimization}], assuming a typical timescale of the evolution given by $\tau = 10 \times 2\pi/J$. Note that the derived configurations are on a proof-of-principle level, i.e., higher fidelity may be still be achievable by further optimization. 

First, we observe that even with a modest number of legs, the transfer fidelity significantly surpasses the maximum population achievable with a single static coupling point in a linear waveguide (which is bounded to $P_{2,\rm max}\simeq 54\%)$, reaching $P_{2,\rm max}=85.9\%$ and $P_{2,\rm max}=89.8\%$ for the four-leg and five-leg configurations, respectively. The coupling strength of each point is around $20\%$ of the hopping rate, yielding $\abs*{g_i^{(1)}}/2\pi\sim 50\,\si{\mega\hertz}$. Such values are routinely achieved and exceeded with current experimental implementations, see, e.g., Refs.~\cite{scigliuzzo_controlling_2022,jouanny_superstrong_2025}.
Second, the average distance between the optimal coupling points in these configurations remains
on the order of the cavity spacing $a$, and therefore does not pose a significant challenge due to the compactness of the CCA~\cite{jouanny_high_2025, jouanny_superstrong_2025}.}

\section{Numerical Optimization}
\label{sec:appendix:optimization}
In the Sec.~\ref{sec:finiteLeg} we optimize the state transfer fidelity for a finite number of legs. Here, we have to impose a constraint for the coupling strengths $\pi\sum_i|g^{(1)}_i|^2/\vg=\tau^{-1}$, based on the decay timescale 
$\tau$ to avoid pathologial results. Setting this constraint to an arbitrary value is justified since rescaling the equations of motion~\eqref{eq:eom} with a parameter $s>0$,
\begin{equation}
x^{(1)}_i\rightarrow x^{(1)}_is, g^{(1)}_i\rightarrow \frac{g_i^{(1)}}{\sqrt{s}}e^{i\kq(x^{(1)}_i-x^{(1)}
_is)}, t\rightarrow ts\,,
\end{equation}
leaves the populations unchanged, but changes the value of the constraint. The rescaling corresponds to reducing the decay rate and increasing the corresponding decay time scale by $s$, while stretching the giant atom to adapt to the extended single photon pulse now emitted upon decay. At the same time the couplings have to be corrected for the changing propagation phases. Note that rescaling of the distance $d\rightarrow ds$ can be omitted as long as emission and absorption are well separated. The scale invariance implies that our optimization results can be mapped to any value of the above constraint.

 \hll{Including the constraint, the cost function of our optimization is $f=-P_\mathrm{2,max}+\beta |C-\tau^{-1}|$, with $C=\pi\sum_i|g_i^{(1)}|^2/\vg$, where we find $\beta=0.1$ to be a suitable weight for all optimizations. We simulate the dynamics using exact diagonalization and optimize the cost function by gradient descent  starting from many random initial conditions. Specifically, we start from closely spaced leg positions (drawn randomly from a Gaussian distribution with a width on the qubit transition wavelength scale), which are then expanded to the pulse-width scale during optimization. Further, we initialize the coupling points with homogeneous strength $1/\sqrt{\tau\pi N}$, which is compatible with the constraint, and random phases. To ensure convergence with respect to the optimization parameters, we sample many different initial conditions. Specifically, for up to 6 legs, we perform 4 optimization runs starting from 500 initial conditions each (amounting to 2000 initial conditions in total). From 7 legs on we perform 8 optimization runs, again with 500 initial conditions each, (amounting to 4000 initial conditions). The increase is necessary as the number of optimization parameters is linear in $N$ and thus hinders convergence. We consider the results converged when the optima from at least two of the independent optimization runs (containing 500 initial conditions) are different only at the $10^{-3}$ scale in the transfer fidelity. }
 
 \hll{Note that we did not constrain the legs to have a minimum distance in our optimization. Nevertheless, we find that the minimal distance between any two legs is not less than about $5\lambda_\mathrm{q}$, with the wavelength of the qubit transition $\lambda_\mathrm{q}=2\pi/\kq$.}
\subsection{Optimization of Pulse Shape}
\label{sec:appendixPulses}
Giant atoms in principle also allow for the engineering of specific emitted pulse shapes. We can compute such designs by minimizing the difference between the Fourier components of the pulse achieved by discrete couplings [see Eq.~\eqref{eq:giantAtomCoupling}] and the desired target pulse, 
\begin{equation}
    \mathcal{F}_{\mathrm{target}} = \min_{\lbrace g^{(1)}_i, x^{(1)}_i\rbrace}\norm{\xi(k)-\xi_\mathrm{target}(k)}_2\,.
    \label{eq:pulseCostfunction}
\end{equation}
To efficiently optimize for a set of discrete spacial couplings, $ g^{(1)}_i, x^{(1)}_i$, we start by considering the resulting photon pulse shape $\xi(k)$, which we can evaluate via the self-energy of the qubit dynamics. For the discretized case of Eq.~\eqref{eq:giantAtomCoupling} the coupling $g_1(k)$
is an analytic function of $k$, and therefore due to chirality also of frequency and the self-energy can be evaluated analytically using the Sokhotski–Plemelj theorem,
\begin{align}
    &\Sigma(\omega+i0^+) =\notag\\ &\quad=
    -\frac{2\pi i}{\vg}\sum_{n,m}g_n^{(1)*}g^{(1)}_me^{i k(\omega)(x^{(1)}_m-x^{(1)}_n)}\Theta_{1/2}(x^{(1)}_m-x^{(1)}_n)\,,
\end{align}
where $\Theta_{1/2}(0)=1/2$ is important because we consider a discrete sum. With this, we have an analytic form of the pulse emitted from our discrete configuration, 
\begin{equation}
    \xi(k) = \frac{\sum_i g_i^{(1)*}e^{-ik x^{(1)}_i}}{\omega(k)-\wq-\Sigma(\omega(k)+i0^+)}\,,
\end{equation}
with the previously derived expression of the self-energy, i.e., we can efficiently evaluate Eq.~\eqref{eq:pulseCostfunction} for a given set of $ x^{(1)}_i, g^{(1)}_i$.
For the optimization it is convenient to rephrase the coupling function as
\begin{align}
    g_1(k) &= \sum_n \left(g^{(1)}_ne^{ik_\mathrm{q}x^{(1)}_n}\right) e^{i\Delta(k) x^{(1)}_n/\vg}\notag\\ &\equiv\sum_n \tilde{g}^{(1)}_n e^{i\Delta(k) x^{(1)}_n/\vg}\,,
\end{align}
and optimize for $\tilde{g}^{(1)}_n$ instead of $g^{(1)}_n$. This allows to completely eliminate the qubit frequency scale from the optimization and promises to avoid highly oscillatory integrals.

\begin{figure}
    \centering
    \includegraphics[]{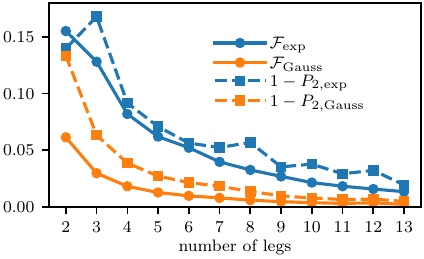}
    \caption{Optimization of giant atoms for particular shapes of emitted pulses. Solid lines display the minimal pulse shape mismatch $\mathcal{F}_\mathrm{target}$ between a target pulse and the pulse emitted by the giant atom as a function of the number of legs. As target pulses, we consider the time-reversal-symmetric wavepackets associated with Eqs.~\eqref{eq:exponential} (blue) and \eqref{eq:Gaussian} (orange). For comparison, we also evaluate the transfer fidelity that can be achieved using the optimized pulses (dashed lines, plotted as $1-P_{2,\mathrm{target}}$).}
    \label{fig:fig2.5}
\end{figure}

Similar optimizations have been performed in~\cite{wang_realizing_2024} towards optimizing for effects of bandstructures, however, with the explicit constraint of leg spacing $\ll \vg\tau$ to avoid retardation effects.  In Fig.~\ref{fig:fig2.5}, we exemplarily perform the optimization for the time-reversal-symmetric pulse shapes of the exponential and Gaussian decays [see Eqs.~\eqref{eq:exponential} and \eqref{eq:Gaussian}]. We find that the Gaussian decay pulse can be approximated significantly better than the exponential decay one. For these optimized results, we also plot the maximal excitation probability $P_2 = \max_t |c_2(t)|^2$ on the right axis of Fig.~\ref{fig:fig2.5}. We find that while for the Gaussian decay, the $P_2$ results align well with the target-pulse overlap, there is significant deviation for pulse from the exponential decay. We can understand this when considering that the key quantity for maximizing $P_2$ is time-reversal symmetry rather than complying with a specific pulse. 

\end{appendix}
\bibliography{references_hardcode}

@ARTICLE{zhang_high-performance_2022,
  title     = "High-performance superconducting quantum processors via laser
               annealing of transmon qubits",
  author    = "Zhang, Eric J and Srinivasan, Srikanth and Sundaresan, Neereja
               and Bogorin, Daniela F and Martin, Yves and Hertzberg, Jared B
               and Timmerwilke, John and Pritchett, Emily J and Yau, Jeng-Bang
               and Wang, Cindy and Landers, William and Lewandowski, Eric P and
               Narasgond, Adinath and Rosenblatt, Sami and Keefe, George A and
               Lauer, Isaac and Rothwell, Mary Beth and McClure, Douglas T and
               Dial, Oliver E and Orcutt, Jason S and Brink, Markus and Chow,
               Jerry M",
  journal   = "Sci. Adv.",
  publisher = "American Association for the Advancement of Science (AAAS)",
  volume    =  8,
  number    =  19,
  pages     = "eabi6690",
  month     =  may,
  year      =  2022,
}

@misc{yang_2026_many,
  title={Many-body quantum optics in a cascaded chiral network}, 
      author={Frank Yang and Parth S. Shah and Chaitali Joshi and Mohammad Mirhosseini},
      year={2026},
      eprint={2607.05760},
      archivePrefix={arXiv},
      primaryClass={quant-ph},
      url={https://arxiv.org/abs/2607.05760}, 
}

@ARTICLE{vepsalainen_improving_2022,
  title     = "Improving qubit coherence using closed-loop feedback",
  author    = "Veps{\"a}l{\"a}inen, Antti and Winik, Roni and Karamlou, Amir H
               and Braum{\"u}ller, Jochen and Paolo, Agustin Di and Sung,
               Youngkyu and Kannan, Bharath and Kjaergaard, Morten and Kim,
               David K and Melville, Alexander J and Niedzielski, Bethany M and
               Yoder, Jonilyn L and Gustavsson, Simon and Oliver, William D",
  journal   = "Nat. Commun.",
  publisher = "Springer Science and Business Media LLC",
  volume    =  13,
  number    =  1,
  pages     = "1932",
  month     =  apr,
  year      =  2022,
  copyright = "https://creativecommons.org/licenses/by/4.0",
}

@article{vieu_electron_2000,
title = {Electron beam lithography: resolution limits and applications},
journal = {Appl. Surf. Sci.},
volume = {164},
number = {1},
pages = {111-117},
year = {2000},
issn = {0169-4332},
doi = {https://doi.org/10.1016/S0169-4332(00)00352-4},
url = {https://www.sciencedirect.com/science/article/pii/S0169433200003524},
author = {C. Vieu and F. Carcenac and A. Pépin and Y. Chen and M. Mejias and A. Lebib and L. Manin-Ferlazzo and L. Couraud and H. Launois}
}

@article{xu_chirality_2026,
	title = {Chirality, nonreciprocity and symmetries for a giant atom},
	volume = {69},
	issn = {1674-7348, 1869-1927},
	url = {https://link.springer.com/10.1007/s11433-025-2854-8},
	doi = {10.1007/s11433-025-2854-8},
	number = {4},
	urldate = {2026-07-13},
	journal = {Sci. China Phys. Mech. Astron.},
	author = {Xu, Luting and Guo, Lingzhen},
	month = apr,
	year = {2026},
	pages = {240311},
}

@article{guo_oscillating_2020,
  title = {Oscillating bound states for a giant atom},
  author = {Guo, Lingzhen and Kockum, Anton Frisk and Marquardt, Florian and Johansson, G\"oran},
  journal = {Phys. Rev. Res.},
  volume = {2},
  issue = {4},
  pages = {043014},
  numpages = {15},
  year = {2020},
  month = {Oct},
  publisher = {American Physical Society},
  doi = {10.1103/PhysRevResearch.2.043014},
  url = {https://link.aps.org/doi/10.1103/PhysRevResearch.2.043014}
}

@article{wang_chiral_2022,
	title = {Chiral quantum network with giant atoms},
	volume = {7},
	issn = {2058-9565},
	url = {https://iopscience.iop.org/article/10.1088/2058-9565/ac6a04},
	doi = {10.1088/2058-9565/ac6a04},
	number = {3},
	urldate = {2025-10-08},
	journal = {Quantum Sci. Technol.},
	author = {Wang, Xin and Li, Hong-Rong},
	month = jul,
	year = {2022},
	pages = {035007},
}

@misc{jouanny_superstrong_2025,
	title = {Superstrong {Dynamics} and {Chiral} {Emission} of a {Giant} {Atom} in a {Structured} {Bath}},
	url = {http://arxiv.org/abs/2509.01579},
	doi = {10.48550/arXiv.2509.01579},
	urldate = {2025-10-16},
	publisher = {arXiv},
	author = {Jouanny, Vincent and Peyruchat, Léo and Scigliuzzo, Marco and Mercurio, Alberto and Benedetto, Enrico Di and Bernardis, Daniele De and Sbroggiò, Davide and Frasca, Simone and Savona, Vincenzo and Ciccarello, Francesco and Scarlino, Pasquale},
	month = sep,
	year = {2025},
	note = {arXiv:2509.01579 [quant-ph]},
}

@article{pechal_microwave-controlled_2014,
	title = {Microwave-{Controlled} {Generation} of {Shaped} {Single} {Photons} in {Circuit} {Quantum} {Electrodynamics}},
	volume = {4},
	copyright = {http://creativecommons.org/licenses/by/3.0/},
	issn = {2160-3308},
	url = {https://link.aps.org/doi/10.1103/PhysRevX.4.041010},
	doi = {10.1103/PhysRevX.4.041010},
	number = {4},
	urldate = {2025-10-24},
	journal = {Phys. Rev. X},
	author = {Pechal, M. and Huthmacher, L. and Eichler, C. and Zeytinoğlu, S. and Abdumalikov, A. A. and Berger, S. and Wallraff, A. and Filipp, S.},
	month = oct,
	year = {2014},
	pages = {041010},
}

@article{miyamura_generation_2025,
	title = {Generation of {Frequency}-{Tunable} {Shaped} {Single} {Microwave} {Photons} {Using} a {Fixed}-{Frequency} {Superconducting} {Qubit}},
	volume = {6},
	issn = {2691-3399},
	url = {https://link.aps.org/doi/10.1103/PRXQuantum.6.020347},
	doi = {10.1103/PRXQuantum.6.020347},
	number = {2},
	urldate = {2025-10-24},
	journal = {PRX Quantum},
	author = {Miyamura, Takeaki and Sunada, Yoshiki and Wang, Zhiling and Ilves, Jesper and Matsuura, Kohei and Nakamura, Yasunobu},
	month = jun,
	year = {2025},
	pages = {020347},
}

@article{kurpiers_deterministic_2018,
	title = {Deterministic quantum state transfer and remote entanglement using microwave photons},
	volume = {558},
	issn = {0028-0836, 1476-4687},
	url = {https://www.nature.com/articles/s41586-018-0195-y},
	doi = {10.1038/s41586-018-0195-y},
	number = {7709},
	urldate = {2025-10-24},
	journal = {Nature},
	author = {Kurpiers, P. and Magnard, P. and Walter, T. and Royer, B. and Pechal, M. and Heinsoo, J. and Salathé, Y. and Akin, A. and Storz, S. and Besse, J.-C. and Gasparinetti, S. and Blais, A. and Wallraff, A.},
	month = jun,
	year = {2018},
	pages = {264--267},
}

@article{axline_-demand_2018,
	title = {On-demand quantum state transfer and entanglement between remote microwave cavity memories},
	volume = {14},
	issn = {1745-2473, 1745-2481},
	url = {https://www.nature.com/articles/s41567-018-0115-y},
	doi = {10.1038/s41567-018-0115-y},
	number = {7},
	urldate = {2025-10-26},
	journal = {Nature Phys},
	author = {Axline, Christopher J. and Burkhart, Luke D. and Pfaff, Wolfgang and Zhang, Mengzhen and Chou, Kevin and Campagne-Ibarcq, Philippe and Reinhold, Philip and Frunzio, Luigi and Girvin, S. M. and Jiang, Liang and Devoret, M. H. and Schoelkopf, R. J.},
	month = jul,
	year = {2018},
	pages = {705--710},
}

@article{gustafsson_propagating_2014,
	title = {Propagating phonons coupled to an artificial atom},
	volume = {346},
	issn = {0036-8075, 1095-9203},
	url = {https://www.science.org/doi/10.1126/science.1257219},
	doi = {10.1126/science.1257219},
	number = {6206},
	urldate = {2025-10-29},
	journal = {Science},
	author = {Gustafsson, Martin V. and Aref, Thomas and Kockum, Anton Frisk and Ekström, Maria K. and Johansson, Göran and Delsing, Per},
	month = oct,
	year = {2014},
	pages = {207--211},
}

@article{bienfait_phonon-mediated_2019,
	title = {Phonon-mediated quantum state transfer and remote qubit entanglement},
	volume = {364},
	issn = {0036-8075, 1095-9203},
	url = {https://www.science.org/doi/10.1126/science.aaw8415},
	doi = {10.1126/science.aaw8415},
	number = {6438},
	urldate = {2025-10-29},
	journal = {Science},
	author = {Bienfait, A. and Satzinger, K. J. and Zhong, Y. P. and Chang, H.-S. and Chou, M.-H. and Conner, C. R. and Dumur, E. and Grebel, J. and Peairs, G. A. and Povey, R. G. and Cleland, A. N.},
	month = apr,
	year = {2019},
	pages = {368--371},
}

@article{dumur_quantum_2021,
	title = {Quantum communication with itinerant surface acoustic wave phonons},
	volume = {7},
	issn = {2056-6387},
	url = {https://www.nature.com/articles/s41534-021-00511-1},
	doi = {10.1038/s41534-021-00511-1},
	number = {1},
	urldate = {2025-10-29},
	journal = {npj Quantum Inf},
	author = {Dumur, E. and Satzinger, K. J. and Peairs, G. A. and Chou, M.-H. and Bienfait, A. and Chang, H.-S. and Conner, C. R. and Grebel, J. and Povey, R. G. and Zhong, Y. P. and Cleland, A. N.},
	month = dec,
	year = {2021},
	pages = {173},
}

@misc{gonzalez-gutierrez_non-markovian_2025,
	title = {Non-{Markovian} dynamics of giant emitters beyond the {Weisskopf}-{Wigner} approximation},
	url = {http://arxiv.org/abs/2508.21784},
	doi = {10.48550/arXiv.2508.21784},
	urldate = {2025-12-03},
	publisher = {arXiv},
	author = {González-Gutiérrez, Carlos A.},
	month = aug,
	year = {2025},
	note = {arXiv:2508.21784 [quant-ph]},
}

@article{wang_realizing_2024,
	title = {Realizing quantum optics in structured environments with giant atoms},
	volume = {6},
	issn = {2643-1564},
	url = {https://link.aps.org/doi/10.1103/PhysRevResearch.6.013279},
	doi = {10.1103/PhysRevResearch.6.013279},
	number = {1},
	urldate = {2025-12-03},
	journal = {Phys. Rev. Research},
	author = {Wang, Xin and Zhu, Huai-Bing and Liu, Tao and Nori, Franco},
	month = mar,
	year = {2024},
	pages = {013279},
}

@article{roccati_controlling_2024,
	title = {Controlling {Markovianity} with {Chiral} {Giant} {Atoms}},
	volume = {133},
	issn = {0031-9007, 1079-7114},
	url = {https://link.aps.org/doi/10.1103/PhysRevLett.133.063603},
	doi = {10.1103/PhysRevLett.133.063603},
	number = {6},
	urldate = {2025-12-04},
	journal = {Phys. Rev. Lett.},
	author = {Roccati, Federico and Cilluffo, Dario},
	month = aug,
	year = {2024},
	pages = {063603},
}

@article{suarez-forero_chiral_2025,
	title = {Chiral {Quantum} {Optics}: {Recent} {Developments} and {Future} {Directions}},
	volume = {6},
	issn = {2691-3399},
	shorttitle = {Chiral {Quantum} {Optics}},
	url = {https://link.aps.org/doi/10.1103/PRXQuantum.6.020101},
	doi = {10.1103/PRXQuantum.6.020101},
	number = {2},
	urldate = {2025-12-04},
	journal = {PRX Quantum},
	author = {Suárez-Forero, D.G. and Jalali Mehrabad, M. and Vega, C. and González-Tudela, A. and Hafezi, M.},
	month = apr,
	year = {2025},
	pages = {020101},
}

@article{andersson_non-exponential_2019,
	title = {Non-exponential decay of a giant artificial atom},
	volume = {15},
	issn = {1745-2473, 1745-2481},
	url = {https://www.nature.com/articles/s41567-019-0605-6},
	doi = {10.1038/s41567-019-0605-6},
	number = {11},
	urldate = {2025-12-04},
	journal = {Nat. Phys.},
	author = {Andersson, Gustav and Suri, Baladitya and Guo, Lingzhen and Aref, Thomas and Delsing, Per},
	month = nov,
	year = {2019},
	pages = {1123--1127},
}

@article{tian_static_2021,
	title = {Static {Hybrid} {Quantum} {Nodes}: {Toward} {Perfect} {State} {Transfer} on a {Photonic} {Chip}},
	volume = {15},
	issn = {2331-7019},
	shorttitle = {Static {Hybrid} {Quantum} {Nodes}},
	url = {https://link.aps.org/doi/10.1103/PhysRevApplied.15.054043},
	doi = {10.1103/PhysRevApplied.15.054043},
	number = {5},
	urldate = {2026-01-12},
	journal = {Phys. Rev. Applied},
	author = {Tian, Zhaohua and Zhang, Pu and Chen, Xue-Wen},
	month = may,
	year = {2021},
	pages = {054043},
}

@article{frisk_kockum_designing_2014,
	title = {Designing frequency-dependent relaxation rates and {Lamb} shifts for a giant artificial atom},
	volume = {90},
	copyright = {http://link.aps.org/licenses/aps-default-license},
	issn = {1050-2947, 1094-1622},
	url = {https://link.aps.org/doi/10.1103/PhysRevA.90.013837},
	doi = {10.1103/PhysRevA.90.013837},
	number = {1},
	urldate = {2026-01-12},
	journal = {Phys. Rev. A},
	author = {Frisk Kockum, Anton and Delsing, Per and Johansson, Göran},
	month = jul,
	year = {2014},
	pages = {013837},
}

@article{vadiraj_engineering_2021,
	title = {Engineering the level structure of a giant artificial atom in waveguide quantum electrodynamics},
	volume = {103},
	issn = {2469-9926, 2469-9934},
	url = {https://link.aps.org/doi/10.1103/PhysRevA.103.023710},
	doi = {10.1103/PhysRevA.103.023710},
	number = {2},
	urldate = {2026-01-22},
	journal = {Phys. Rev. A},
	author = {Vadiraj, A. M. and Ask, Andreas and McConkey, T. G. and Nsanzineza, I. and Chang, C. W. Sandbo and Kockum, Anton Frisk and Wilson, C. M.},
	month = feb,
	year = {2021},
	pages = {023710},
}

@article{guo_giant_2017,
	title = {Giant acoustic atom: {A} single quantum system with a deterministic time delay},
	volume = {95},
	copyright = {http://link.aps.org/licenses/aps-default-license},
	issn = {2469-9926, 2469-9934},
	shorttitle = {Giant acoustic atom},
	url = {http://link.aps.org/doi/10.1103/PhysRevA.95.053821},
	doi = {10.1103/PhysRevA.95.053821},
	number = {5},
	urldate = {2026-01-23},
	journal = {Phys. Rev. A},
	author = {Guo, Lingzhen and Grimsmo, Arne and Kockum, Anton Frisk and Pletyukhov, Mikhail and Johansson, Göran},
	month = may,
	year = {2017},
	pages = {053821},
}

@article{kockum_decoherence-free_2018,
	title = {Decoherence-{Free} {Interaction} between {Giant} {Atoms} in {Waveguide} {Quantum} {Electrodynamics}},
	volume = {120},
	issn = {0031-9007, 1079-7114},
	url = {https://link.aps.org/doi/10.1103/PhysRevLett.120.140404},
	doi = {10.1103/PhysRevLett.120.140404},
	number = {14},
	urldate = {2026-01-23},
	journal = {Phys. Rev. Lett.},
	author = {Kockum, Anton Frisk and Johansson, Göran and Nori, Franco},
	month = apr,
	year = {2018},
	pages = {140404},
}

@incollection{hadfield_quantum_2016,
	address = {Cham},
	title = {Quantum {Acoustics} with {Surface} {Acoustic} {Waves}},
	url = {http://link.springer.com/10.1007/978-3-319-24091-6_9},
	doi = {10.1007/978-3-319-24091-6_9},
	urldate = {2026-01-29},
	booktitle = {Superconducting {Devices} in {Quantum} {Optics}},
	publisher = {Springer International Publishing},
	author = {Aref, Thomas and Delsing, Per and Ekström, Maria K. and Kockum, Anton Frisk and Gustafsson, Martin V. and Johansson, Göran and Leek, Peter J. and Magnusson, Einar and Manenti, Riccardo},
	editor = {Hadfield, Robert H. and Johansson, Göran},
	year = {2016},
	note = {Series Title: Quantum Science and Technology},
	pages = {217--244},
}

@article{soro_chiral_2022,
	title = {Chiral quantum optics with giant atoms},
	volume = {105},
	issn = {2469-9926, 2469-9934},
	url = {https://link.aps.org/doi/10.1103/PhysRevA.105.023712},
	doi = {10.1103/PhysRevA.105.023712},
	number = {2},
	urldate = {2026-02-02},
	journal = {Phys. Rev. A},
	author = {Soro, Ariadna and Kockum, Anton Frisk},
	month = feb,
	year = {2022},
	pages = {023712},
}

@article{pernas_shaping_2026,
	title     = "Shaping frequency-tunable single photons for quantum networking
               in waveguide {QED}",
  author    = "Pernas, {\'A}lvaro and Puebla, Ricardo",
  journal   = "Quantum Sci. Technol.",
  publisher = "IOP Publishing",
  volume    =  11,
  number    =  4,
  pages     = "045027",
  month     =  dec,
  year      =  2026,
  copyright = "https://publishingsupport.iopscience.iop.org/iop-standard/v1"
}

@article{magnard_microwave_2020,
	title = {Microwave {Quantum} {Link} between {Superconducting} {Circuits} {Housed} in {Spatially} {Separated} {Cryogenic} {Systems}},
	volume = {125},
	issn = {0031-9007, 1079-7114},
	url = {https://link.aps.org/doi/10.1103/PhysRevLett.125.260502},
	doi = {10.1103/PhysRevLett.125.260502},
	number = {26},
	urldate = {2026-03-10},
	journal = {Phys. Rev. Lett.},
	author = {Magnard, P. and Storz, S. and Kurpiers, P. and Schär, J. and Marxer, F. and Lütolf, J. and Walter, T. and Besse, J.-C. and Gabureac, M. and Reuer, K. and Akin, A. and Royer, B. and Blais, A. and Wallraff, A.},
	month = dec,
	year = {2020},
	pages = {260502},
}

@article{mirhosseini_superconducting_2018,
	title = {Superconducting metamaterials for waveguide quantum electrodynamics},
	volume = {9},
	issn = {2041-1723},
	url = {https://www.nature.com/articles/s41467-018-06142-z},
	doi = {10.1038/s41467-018-06142-z},
	number = {1},
	urldate = {2026-03-16},
	journal = {Nat Commun},
	author = {Mirhosseini, Mohammad and Kim, Eunjong and Ferreira, Vinicius S. and Kalaee, Mahmoud and Sipahigil, Alp and Keller, Andrew J. and Painter, Oskar},
	month = sep,
	year = {2018},
	pages = {3706},
}

@article{lin_giant_2026,
	title = {Giant atoms coupled to a waveguide: Continuous coupling and multiple excitations},
  author = {Lin, Shiying and Zhao, Xinyu and Xia, Yan},
  journal = {Phys. Rev. A},
  volume = {113},
  issue = {3},
  pages = {033705},
  numpages = {18},
  year = {2026},
  month = {Mar},
  publisher = {American Physical Society},
  doi = {10.1103/1d3w-g932},
  url = {https://link.aps.org/doi/10.1103/1d3w-g932}
}

@article{wang_efficient_2011,
	title = {Efficient excitation of a two-level atom by a single photon in a propagating mode},
	volume = {83},
	copyright = {http://link.aps.org/licenses/aps-default-license},
	issn = {1050-2947, 1094-1622},
	url = {https://link.aps.org/doi/10.1103/PhysRevA.83.063842},
	doi = {10.1103/PhysRevA.83.063842},
	number = {6},
	urldate = {2026-03-16},
	journal = {Phys. Rev. A},
	author = {Wang, Yimin and Minář, Jiří and Sheridan, Lana and Scarani, Valerio},
	month = jun,
	year = {2011},
	pages = {063842},
}

@article{vogell_deterministic_2017,
	title = {Deterministic quantum state transfer between remote qubits in cavities},
	volume = {2},
	issn = {2058-9565},
	url = {https://iopscience.iop.org/article/10.1088/2058-9565/aa868b},
	doi = {10.1088/2058-9565/aa868b},
	number = {4},
	urldate = {2026-03-16},
	journal = {Quantum Sci. Technol.},
	author = {Vogell, B and Vermersch, B and Northup, T E and Lanyon, B P and Muschik, C A},
	month = dec,
	year = {2017},
	pages = {045003},
}

@article{morin_deterministic_2019,
	title = {Deterministic {Shaping} and {Reshaping} of {Single}-{Photon} {Temporal} {Wave} {Functions}},
	volume = {123},
	issn = {0031-9007, 1079-7114},
	url = {https://link.aps.org/doi/10.1103/PhysRevLett.123.133602},
	doi = {10.1103/PhysRevLett.123.133602},
	number = {13},
	urldate = {2026-03-16},
	journal = {Phys. Rev. Lett.},
	author = {Morin, O. and Körber, M. and Langenfeld, S. and Rempe, G.},
	month = sep,
	year = {2019},
	pages = {133602},
}

@article{carollo_mechanism_2020,
	title = {Mechanism of decoherence-free coupling between giant atoms},
	volume = {2},
	issn = {2643-1564},
	url = {https://link.aps.org/doi/10.1103/PhysRevResearch.2.043184},
	doi = {10.1103/PhysRevResearch.2.043184},
	number = {4},
	urldate = {2026-03-16},
	journal = {Phys. Rev. Research},
	author = {Carollo, Angelo and Cilluffo, Dario and Ciccarello, Francesco},
	month = nov,
	year = {2020},
	pages = {043184},
}

@article{bose_quantum_2003,
	title = {Quantum {Communication} through an {Unmodulated} {Spin} {Chain}},
	volume = {91},
	copyright = {http://link.aps.org/licenses/aps-default-license},
	issn = {0031-9007, 1079-7114},
	url = {https://link.aps.org/doi/10.1103/PhysRevLett.91.207901},
	doi = {10.1103/PhysRevLett.91.207901},
	number = {20},
	urldate = {2026-03-17},
	journal = {Phys. Rev. Lett.},
	author = {Bose, Sougato},
	month = nov,
	year = {2003},
	pages = {207901},
}

@article{banchi_ballistic_2013,
	title = {Ballistic quantum state transfer in spin chains: {General} theory for quasi-free models and arbitrary initial states},
	volume = {128},
	copyright = {http://www.springer.com/tdm},
	issn = {2190-5444},
	shorttitle = {Ballistic quantum state transfer in spin chains},
	url = {http://link.springer.com/10.1140/epjp/i2013-13137-6},
	doi = {10.1140/epjp/i2013-13137-6},
	number = {11},
	urldate = {2026-03-17},
	journal = {Eur. Phys. J. Plus},
	author = {Banchi, Leonardo},
	month = nov,
	year = {2013},
	pages = {137},
}

@misc{bernard_distinctive_2024,
	title = {Distinctive features of inhomogeneous spin chains},
	copyright = {arXiv.org perpetual, non-exclusive license},
	url = {https://arxiv.org/abs/2411.09487},
	doi = {10.48550/ARXIV.2411.09487},
	urldate = {2026-03-17},
	publisher = {arXiv},
	author = {Bernard, Pierre-Antoine and Parez, Gilles and Vinet, Luc},
	year = {2024},
	note = {arXiv:2411.09487 [quant-ph]},
}

@article{krantz_quantum_2019,
	title = {A quantum engineer's guide to superconducting qubits},
	volume = {6},
	issn = {1931-9401},
	url = {https://pubs.aip.org/apr/article/6/2/021318/570326/A-quantum-engineer-s-guide-to-superconducting},
	doi = {10.1063/1.5089550},
	number = {2},
	urldate = {2026-03-19},
	journal = {Applied Physics Reviews},
	author = {Krantz, P. and Kjaergaard, M. and Yan, F. and Orlando, T. P. and Gustavsson, S. and Oliver, W. D.},
	month = jun,
	year = {2019},
	pages = {021318},
}

@article{bradac_quantum_2019,
	title = {Quantum nanophotonics with group {IV} defects in diamond},
	volume = {10},
	issn = {2041-1723},
	url = {https://www.nature.com/articles/s41467-019-13332-w},
	doi = {10.1038/s41467-019-13332-w},
	number = {1},
	urldate = {2026-03-19},
	journal = {Nat Commun},
	author = {Bradac, Carlo and Gao, Weibo and Forneris, Jacopo and Trusheim, Matthew E. and Aharonovich, Igor},
	month = dec,
	year = {2019},
	pages = {5625},
}

@article{du_single-photon_2021,
	title = {Single-photon nonreciprocal excitation transfer with non-{Markovian} retarded effects},
	volume = {103},
	issn = {2469-9926, 2469-9934},
	url = {https://link.aps.org/doi/10.1103/PhysRevA.103.053701},
	doi = {10.1103/PhysRevA.103.053701},
	number = {5},
	urldate = {2026-03-19},
	journal = {Phys. Rev. A},
	author = {Du, Lei and Cai, Mao-Rui and Wu, Jin-Hui and Wang, Zhihai and Li, Yong},
	month = may,
	year = {2021},
	pages = {053701},
}

@article{xiao_bound_2022,
	title = {Bound state in a giant atom-modulated resonators system},
	volume = {8},
	issn = {2056-6387},
	url = {https://www.nature.com/articles/s41534-022-00591-7},
	doi = {10.1038/s41534-022-00591-7},
	number = {1},
	urldate = {2026-03-19},
	journal = {npj Quantum Inf},
	author = {Xiao, Han and Wang, Luojia and Li, Zheng-Hong and Chen, Xianfeng and Yuan, Luqi},
	month = jul,
	year = {2022},
	pages = {80},
}

@article{longhi_photonic_2020,
	title = {Photonic simulation of giant atom decay},
	volume = {45},
	issn = {0146-9592, 1539-4794},
	url = {https://opg.optica.org/abstract.cfm?URI=ol-45-11-3017},
	doi = {10.1364/OL.393578},
	number = {11},
	urldate = {2026-03-19},
	journal = {Opt. Lett.},
	author = {Longhi, Stefano},
	month = jun,
	year = {2020},
	pages = {3017},
}

@article{yin_non-markovian_2022,
	title = {Non-{Markovian} disentanglement dynamics in double-giant-atom waveguide-{QED} systems},
	volume = {106},
	issn = {2469-9926, 2469-9934},
	url = {https://link.aps.org/doi/10.1103/PhysRevA.106.063703},
	doi = {10.1103/PhysRevA.106.063703},
	number = {6},
	urldate = {2026-03-19},
	journal = {Phys. Rev. A},
	author = {Yin, Xian-Li and Luo, Wen-Bin and Liao, Jie-Qiao},
	month = dec,
	year = {2022},
	pages = {063703},
}

@article{lim_oscillating_2023,
	title = {Oscillating bound states in non-{Markovian} photonic lattices},
	volume = {107},
	issn = {2469-9926, 2469-9934},
	url = {https://link.aps.org/doi/10.1103/PhysRevA.107.023716},
	doi = {10.1103/PhysRevA.107.023716},
	number = {2},
	urldate = {2026-03-19},
	journal = {Phys. Rev. A},
	author = {Lim, Kian Hwee and Mok, Wai-Keong and Kwek, Leong-Chuan},
	month = feb,
	year = {2023},
	pages = {023716},
}

@article{chen_giant-atom_2023,
	title = {Giant-atom effects on population and entanglement dynamics of {Rydberg} atoms in the optical regime},
	volume = {5},
	issn = {2643-1564},
	url = {https://link.aps.org/doi/10.1103/PhysRevResearch.5.043135},
	doi = {10.1103/PhysRevResearch.5.043135},
	number = {4},
	urldate = {2026-03-19},
	journal = {Phys. Rev. Research},
	author = {Chen, Yao-Tong and Du, Lei and Zhang, Yan and Guo, Lingzhen and Wu, Jin-Hui and Artoni, M. and La Rocca, G. C.},
	month = nov,
	year = {2023},
	pages = {043135},
}

@article{gonzalez-tudela_engineering_2019,
	title = {Engineering and {Harnessing} {Giant} {Atoms} in {High}-{Dimensional} {Baths}: {A} {Proposal} for {Implementation} with {Cold} {Atoms}},
	volume = {122},
	issn = {0031-9007, 1079-7114},
	shorttitle = {Engineering and {Harnessing} {Giant} {Atoms} in {High}-{Dimensional} {Baths}},
	url = {https://link.aps.org/doi/10.1103/PhysRevLett.122.203603},
	doi = {10.1103/PhysRevLett.122.203603},
	number = {20},
	urldate = {2026-03-19},
	journal = {Phys. Rev. Lett.},
	author = {González-Tudela, A. and Muñoz, C. Sánchez and Cirac, J. I.},
	month = may,
	year = {2019},
	pages = {203603},
}

@article{joshi_resonance_2023,
	title = {Resonance {Fluorescence} of a {Chiral} {Artificial} {Atom}},
	volume = {13},
	issn = {2160-3308},
	url = {https://link.aps.org/doi/10.1103/PhysRevX.13.021039},
	doi = {10.1103/PhysRevX.13.021039},
	number = {2},
	urldate = {2026-03-19},
	journal = {Phys. Rev. X},
	author = {Joshi, Chaitali and Yang, Frank and Mirhosseini, Mohammad},
	month = jun,
	year = {2023},
	pages = {021039},
}

@article{leonforte_quantum_2025,
	title = {Quantum optics with giant atoms in a structured photonic bath},
	volume = {10},
	issn = {2058-9565},
	url = {https://iopscience.iop.org/article/10.1088/2058-9565/ada08d},
	doi = {10.1088/2058-9565/ada08d},
	number = {1},
	urldate = {2026-03-19},
	journal = {Quantum Sci. Technol.},
	author = {Leonforte, Luca and Sun, Xuejian and Valenti, Davide and Spagnolo, Bernardo and Illuminati, Fabrizio and Carollo, Angelo and Ciccarello, Francesco},
	month = jan,
	year = {2025},
	pages = {015057},
}

@article{qiu_collective_2023,
	title = {Collective radiance of giant atoms in non-{Markovian} regime},
	volume = {66},
	issn = {1674-7348, 1869-1927},
	url = {https://link.springer.com/10.1007/s11433-022-1990-x},
	doi = {10.1007/s11433-022-1990-x},
	number = {2},
	urldate = {2026-03-19},
	journal = {Sci. China Phys. Mech. Astron.},
	author = {Qiu, Qing-Yang and Wu, Ying and Lü, Xin-You},
	month = feb,
	year = {2023},
	pages = {224212},
}

@article{vega_qubit-photon_2021,
	title = {Qubit-photon bound states in topological waveguides with long-range hoppings},
	volume = {104},
	issn = {2469-9926, 2469-9934},
	url = {https://link.aps.org/doi/10.1103/PhysRevA.104.053522},
	doi = {10.1103/PhysRevA.104.053522},
	number = {5},
	urldate = {2026-03-19},
	journal = {Phys. Rev. A},
	author = {Vega, C. and Bello, M. and Porras, D. and González-Tudela, A.},
	month = nov,
	year = {2021},
	pages = {053522},
}

@article{wang_tunable_2021,
	title = {Tunable {Chiral} {Bound} {States} with {Giant} {Atoms}},
	volume = {126},
	issn = {0031-9007, 1079-7114},
	url = {https://link.aps.org/doi/10.1103/PhysRevLett.126.043602},
	doi = {10.1103/PhysRevLett.126.043602},
	number = {4},
	urldate = {2026-03-19},
	journal = {Phys. Rev. Lett.},
	author = {Wang, Xin and Liu, Tao and Kockum, Anton Frisk and Li, Hong-Rong and Nori, Franco},
	month = jan,
	year = {2021},
	pages = {043602},
}

@article{yu_entanglement_2021,
	title = {Entanglement preparation and nonreciprocal excitation evolution in giant atoms by controllable dissipation and coupling},
	volume = {104},
	issn = {2469-9926, 2469-9934},
	url = {https://link.aps.org/doi/10.1103/PhysRevA.104.013720},
	doi = {10.1103/PhysRevA.104.013720},
	number = {1},
	urldate = {2026-03-19},
	journal = {Phys. Rev. A},
	author = {Yu, Hongwei and Wang, Zhihai and Wu, Jin-Hui},
	month = jul,
	year = {2021},
	pages = {013720},
}

@article{chen_simulating_2025,
	title = {Simulating open quantum systems with giant atoms},
	volume = {10},
	issn = {2058-9565},
	url = {https://iopscience.iop.org/article/10.1088/2058-9565/adb2bd},
	doi = {10.1088/2058-9565/adb2bd},
	number = {2},
	urldate = {2026-03-20},
	journal = {Quantum Sci. Technol.},
	author = {Chen, Guangze and Frisk Kockum, Anton},
	month = apr,
	year = {2025},
	pages = {025028},
}

@article{levy-yeyati_engineering_2026,
	 title = {Engineering giant transmon molecules as mediators of conditional two-photon gates},
  author = {Levy-Yeyati, Tom\'as and Ramos, Tom\'as and Gonz\'alez-Tudela, Alejandro},
  journal = {Phys. Rev. Appl.},
  volume = {25},
  issue = {5},
  pages = {054060},
  numpages = {10},
  year = {2026},
  month = {May},
  publisher = {American Physical Society},
  doi = {10.1103/kv28-r6w4},
  url = {https://link.aps.org/doi/10.1103/kv28-r6w4}
}

@misc{jiang_photon-echo_2026,
	title = {Photon-echo synchronization and quantum state transfer in short quantum links},
	url = {http://arxiv.org/abs/2603.19064},
	doi = {10.48550/arXiv.2603.19064},
	urldate = {2026-03-24},
	publisher = {arXiv},
	author = {Jiang, Hong and Barahona-Pascual, Carlos and García-Ripoll, Juan José},
	month = mar,
	year = {2026},
	note = {arXiv:2603.19064 [quant-ph]},
}

@article{he_quantum_2025,
	title = {Quantum {State} {Transfer} via a {Multimode} {Resonator}},
	volume = {134},
	issn = {0031-9007, 1079-7114},
	url = {https://link.aps.org/doi/10.1103/PhysRevLett.134.023602},
	doi = {10.1103/PhysRevLett.134.023602},
	number = {2},
	urldate = {2026-03-24},
	journal = {Phys. Rev. Lett.},
	author = {He, Yang and Zhang, Yu-Xiang},
	month = jan,
	year = {2025},
	pages = {023602},
}

@article{diekmann_ultrafast_2024,
	title = {Ultrafast {Excitation} {Exchange} in a {Maxwell} {Fish}-{Eye} {Lens}},
	volume = {132},
	url = {https://link.aps.org/doi/10.1103/PhysRevLett.132.013602},
	doi = {10.1103/PhysRevLett.132.013602},
	number = {1},
	journal = {Phys. Rev. Lett.},
	publisher = {American Physical Society},
	author = {Diekmann, Oliver and Krimer, Dmitry O. and Rotter, Stefan},
	month = jan,
	year = {2024},
	pages = {013602},
}

@article{ferreira_collapse_2021,
	title = {Collapse and {Revival} of an {Artificial} {Atom} {Coupled} to a {Structured} {Photonic} {Reservoir}},
	volume = {11},
	issn = {2160-3308},
	url = {https://link.aps.org/doi/10.1103/PhysRevX.11.041043},
	doi = {10.1103/PhysRevX.11.041043},
	number = {4},
	urldate = {2022-09-05},
	journal = {Physical Review X},
	author = {Ferreira, Vinicius S. and Banker, Jash and Sipahigil, Alp and Matheny, Matthew H. and Keller, Andrew J. and Kim, Eunjong and Mirhosseini, Mohammad and Painter, Oskar},
	month = dec,
	year = {2021},
	pages = {041043},
}

@incollection{frisk_kockum_quantum_2021,
	address = {Singapore},
	title = {Quantum {Optics} with {Giant} {Atoms}—the {First} {Five} {Years}},
	volume = {33},
	isbn = {978-981-15-5190-1 978-981-15-5191-8},
	url = {http://link.springer.com/10.1007/978-981-15-5191-8_12},
	doi = {10.1007/978-981-15-5191-8_12},
	urldate = {2025-10-16},
	booktitle = {International {Symposium} on {Mathematics}, {Quantum} {Theory}, and {Cryptography}},
	publisher = {Springer Singapore},
	author = {Frisk Kockum, Anton},
	editor = {Takagi, Tsuyoshi and Wakayama, Masato and Tanaka, Keisuke and Kunihiro, Noboru and Kimoto, Kazufumi and Ikematsu, Yasuhiko},
	year = {2021},
	pages = {125--146},
}

@article{guimond_unidirectional_2020,
	title = {A unidirectional on-chip photonic interface for superconducting circuits},
	volume = {6},
	issn = {2056-6387},
	url = {https://www.nature.com/articles/s41534-020-0261-9},
	doi = {10.1038/s41534-020-0261-9},
	number = {1},
	urldate = {2025-10-16},
	journal = {npj Quantum Inf},
	author = {Guimond, P.-O. and Vermersch, B. and Juan, M. L. and Sharafiev, A. and Kirchmair, G. and Zoller, P.},
	month = mar,
	year = {2020},
	pages = {32},
}

@article{guo_beyond_2020,
	title = {Beyond spontaneous emission: {Giant} atom bounded in the continuum},
	volume = {102},
	issn = {2469-9926, 2469-9934},
	shorttitle = {Beyond spontaneous emission},
	url = {https://link.aps.org/doi/10.1103/PhysRevA.102.033706},
	doi = {10.1103/PhysRevA.102.033706},
	number = {3},
	urldate = {2025-10-22},
	journal = {Phys. Rev. A},
	author = {Guo, Shangjie and Wang, Yidan and Purdy, Thomas and Taylor, Jacob},
	month = sep,
	year = {2020},
	pages = {033706},
}

@article{kannan_waveguide_2020,
	title = {Waveguide quantum electrodynamics with superconducting artificial giant atoms},
	volume = {583},
	issn = {0028-0836, 1476-4687},
	url = {https://www.nature.com/articles/s41586-020-2529-9},
	doi = {10.1038/s41586-020-2529-9},
	number = {7818},
	urldate = {2025-10-22},
	journal = {Nature},
	author = {Kannan, Bharath and Ruckriegel, Max J. and Campbell, Daniel L. and Frisk Kockum, Anton and Braumüller, Jochen and Kim, David K. and Kjaergaard, Morten and Krantz, Philip and Melville, Alexander and Niedzielski, Bethany M. and Vepsäläinen, Antti and Winik, Roni and Yoder, Jonilyn L. and Nori, Franco and Orlando, Terry P. and Gustavsson, Simon and Oliver, William D.},
	month = jul,
	year = {2020},
	pages = {775--779},
}

@article{wang_giant_2022,
	title = {Giant spin ensembles in waveguide magnonics},
	volume = {13},
	issn = {2041-1723},
	url = {https://www.nature.com/articles/s41467-022-35174-9},
	doi = {10.1038/s41467-022-35174-9},
	number = {1},
	urldate = {2025-10-22},
	journal = {Nat Commun},
	author = {Wang, Zi-Qi and Wang, Yi-Pu and Yao, Jiguang and Shen, Rui-Chang and Wu, Wei-Jiang and Qian, Jie and Li, Jie and Zhu, Shi-Yao and You, J. Q.},
	month = dec,
	year = {2022},
	pages = {7580},
}

@article{cirac_quantum_1997,
	title = {Quantum {State} {Transfer} and {Entanglement} {Distribution} among {Distant} {Nodes} in a {Quantum} {Network}},
	volume = {78},
	url = {https://link.aps.org/doi/10.1103/PhysRevLett.78.3221},
	doi = {10.1103/PhysRevLett.78.3221},
	number = {16},
	urldate = {2025-12-18},
	journal = {Phys. Rev. Lett.},
	author = {Cirac, J. I. and Zoller, P. and Kimble, H. J. and Mabuchi, H.},
	month = apr,
	year = {1997},
	pages = {3221--3224},
}

@article{kimble_quantum_2008,
	title = {The quantum internet},
	volume = {453},
	copyright = {2008 Springer Nature Limited},
	issn = {1476-4687},
	url = {https://www.nature.com/articles/nature07127},
	doi = {10.1038/nature07127},
	number = {7198},
	urldate = {2025-12-18},
	journal = {Nature},
	author = {Kimble, H. J.},
	month = jun,
	year = {2008},
	pages = {1023--1030},
}

@article{de_bernardis_chiral_2023,
	title = {Chiral {Quantum} {Optics} in the {Bulk} of {Photonic} {Quantum} {Hall} {Systems}},
	volume = {4},
	url = {https://link.aps.org/doi/10.1103/PRXQuantum.4.030306},
	doi = {10.1103/PRXQuantum.4.030306},
	number = {3},
	urldate = {2025-12-18},
	journal = {PRX Quantum},
	author = {De Bernardis, Daniele and Piccioli, Francesco S. and Rabl, Peter and Carusotto, Iacopo},
	month = jul,
	year = {2023},
	pages = {030306},
}

@article{stobinska_perfect_2009,
	title = {Perfect excitation of a matter qubit by a single photon in free space},
	volume = {86},
	issn = {0295-5075},
	url = {https://doi.org/10.1209/0295-5075/86/14007},
	doi = {10.1209/0295-5075/86/14007},
	number = {1},
	urldate = {2025-12-18},
	journal = {EPL},
	author = {Stobińska, M. and Alber, G. and Leuchs, G.},
	month = apr,
	year = {2009},
	pages = {14007},
}

@article{ritter_elementary_2012,
	title = {An elementary quantum network of single atoms in optical cavities},
	volume = {484},
	copyright = {2012 Springer Nature Limited},
	issn = {1476-4687},
	url = {https://www.nature.com/articles/nature11023},
	doi = {10.1038/nature11023},
	number = {7393},
	urldate = {2025-12-18},
	journal = {Nature},
	author = {Ritter, Stephan and Nölleke, Christian and Hahn, Carolin and Reiserer, Andreas and Neuzner, Andreas and Uphoff, Manuel and Mücke, Martin and Figueroa, Eden and Bochmann, Joerg and Rempe, Gerhard},
	month = apr,
	year = {2012},
	pages = {195--200},
}

@article{penas_improving_2023,
	title = {Improving quantum state transfer: correcting non-{Markovian} and distortion effects},
	volume = {8},
	issn = {2058-9565},
	shorttitle = {Improving quantum state transfer},
	url = {https://doi.org/10.1088/2058-9565/acf60a},
	doi = {10.1088/2058-9565/acf60a},
	number = {4},
	urldate = {2025-12-18},
	journal = {Quantum Sci. Technol.},
	author = {Peñas, Guillermo F and Puebla, Ricardo and José García-Ripoll, Juan},
	month = sep,
	year = {2023},
	pages = {045026},
}

@article{penas_multiplexed_2024,
	title = {Multiplexed quantum state transfer in waveguides},
	volume = {6},
	url = {https://link.aps.org/doi/10.1103/PhysRevResearch.6.033294},
	doi = {10.1103/PhysRevResearch.6.033294},
	number = {3},
	urldate = {2025-12-18},
	journal = {Phys. Rev. Res.},
	author = {Peñas, Guillermo F. and Puebla, Ricardo and García-Ripoll, Juan José},
	month = sep,
	year = {2024},
	pages = {033294},
}

@article{lodahl_interfacing_2015,
	title = {Interfacing single photons and single quantum dots with photonic nanostructures},
	volume = {87},
	url = {https://link.aps.org/doi/10.1103/RevModPhys.87.347},
	doi = {10.1103/RevModPhys.87.347},
	number = {2},
	urldate = {2025-12-18},
	journal = {Rev. Mod. Phys.},
	author = {Lodahl, Peter and Mahmoodian, Sahand and Stobbe, Sørenglish},
	month = may,
	year = {2015},
	pages = {347--400},
}

@article{ramos_non-markovian_2016,
	title = {Non-{Markovian} dynamics in chiral quantum networks with spins and photons},
	volume = {93},
	url = {https://link.aps.org/doi/10.1103/PhysRevA.93.062104},
	doi = {10.1103/PhysRevA.93.062104},
	number = {6},
	urldate = {2025-12-18},
	journal = {Phys. Rev. A},
	author = {Ramos, Tomás and Vermersch, Benoît and Hauke, Philipp and Pichler, Hannes and Zoller, Peter},
	month = jun,
	year = {2016},
	pages = {062104},
}

@book{cohen-tannoudji_atom-photon_2008,
	address = {New York, NY},
	series = {A {Wiley}-{Interscience} publication},
	title = {Atom-photon interactions: basic processes and applications},
	shorttitle = {Atom-photon interactions},
	doi = {10.1002/9783527617197},
	publisher = {Wiley},
	editor = {Cohen-Tannoudji, Claude and Dupont-Roc, Jacques and Grynberg, Gilbert},
	year = {2008},
}

@article{gonzalez-tudela_markovian_2017,
	title = {Markovian and non-{Markovian} dynamics of quantum emitters coupled to two-dimensional structured reservoirs},
	volume = {96},
	url = {https://link.aps.org/doi/10.1103/PhysRevA.96.043811},
	doi = {10.1103/PhysRevA.96.043811},
	number = {4},
	urldate = {2025-12-18},
	journal = {Phys. Rev. A},
	author = {González-Tudela, A. and Cirac, J. I.},
	month = oct,
	year = {2017},
	pages = {043811},
}

@misc{wang_decoherence-free_2026,
	title = {Decoherence-{Free} {Qubit} and {Chiral} {Emission} from a {Giant} {Molecule} in {Waveguide} {QED}},
	url = {http://arxiv.org/abs/2603.27443},
	doi = {10.48550/arXiv.2603.27443},
	urldate = {2026-04-05},
	publisher = {arXiv},
	author = {Wang, Yang and García-Ripoll, Juan José and Santos, Alan C.},
	month = mar,
	year = {2026},
	note = {arXiv:2603.27443 [quant-ph]},
}

@misc{chen_efficient_2026,
	title = {Efficient three-qubit gates with giant atoms},
	url = {http://arxiv.org/abs/2510.04545},
	doi = {10.48550/arXiv.2510.04545},
	urldate = {2026-04-05},
	publisher = {arXiv},
	author = {Chen, Guangze and Kockum, Anton Frisk},
	month = jan,
	year = {2026},
	note = {arXiv:2510.04545 [quant-ph]},
}

@article{han_giant_2026,
	title = {Giant atom with disorders: {Effects} from imperfect couplings},
	volume = {113},
	issn = {2469-9926, 2469-9934},
	shorttitle = {Giant atom with disorders},
	url = {https://link.aps.org/doi/10.1103/1px2-2db4},
	doi = {10.1103/1px2-2db4},
	number = {1},
	urldate = {2026-04-05},
	journal = {Phys. Rev. A},
	author = {Han, Muming and Guo, Lingzhen},
	month = jan,
	year = {2026},
	pages = {013716},
}

@article{xu_catch_2024,
	title = {Catch and release of propagating bosonic field with non-{Markovian} giant atom},
	volume = {26},
	issn = {1367-2630},
	url = {https://iopscience.iop.org/article/10.1088/1367-2630/ad18ed},
	doi = {10.1088/1367-2630/ad18ed},
	number = {1},
	urldate = {2026-04-05},
	journal = {New J. Phys.},
	author = {Xu, Luting and Guo, Lingzhen},
	month = jan,
	year = {2024},
	pages = {013025},
}

@misc{rieck_controlled-z_2026,
	title = {Controlled-{Z} gates with giant atoms in structured waveguides},
	url = {http://arxiv.org/abs/2603.26345},
	doi = {10.48550/arXiv.2603.26345},
	urldate = {2026-04-10},
	publisher = {arXiv},
	author = {Rieck, Walter and Soro, Ariadna and Kockum, Anton Frisk and Chen, Guangze},
	month = mar,
	year = {2026},
	note = {arXiv:2603.26345 [quant-ph]},
}

@article{casulleras_remote_2021,
	title = {Remote {Individual} {Addressing} of {Quantum} {Emitters} with {Chirped} {Pulses}},
	volume = {126},
	issn = {0031-9007, 1079-7114},
	url = {https://link.aps.org/doi/10.1103/PhysRevLett.126.103602},
	doi = {10.1103/PhysRevLett.126.103602},
	number = {10},
	urldate = {2026-04-13},
	journal = {Phys. Rev. Lett.},
	author = {Casulleras, S. and Gonzalez-Ballestero, C. and Maurer, P. and García-Ripoll, J. J. and Romero-Isart, O.},
	month = mar,
	year = {2021},
	pages = {103602},
}

@article{jouanny_high_2025,
	title = {High kinetic inductance cavity arrays for compact band engineering and topology-based disorder meters},
	volume = {16},
	issn = {2041-1723},
	url = {https://www.nature.com/articles/s41467-025-58595-8},
	doi = {10.1038/s41467-025-58595-8},
	number = {1},
	urldate = {2026-04-15},
	journal = {Nat Commun},
	author = {Jouanny, Vincent and Frasca, Simone and Weibel, Vera Jo and Peyruchat, Léo and Scigliuzzo, Marco and Oppliger, Fabian and De Palma, Franco and Sbroggiò, Davide and Beaulieu, Guillaume and Zilberberg, Oded and Scarlino, Pasquale},
	month = apr,
	year = {2025},
	pages = {3396},
}

@article{scigliuzzo_controlling_2022,
  title = {Controlling Atom-Photon Bound States in an Array of Josephson-Junction Resonators},
  author = {Scigliuzzo, Marco and Calaj\`o, Giuseppe and Ciccarello, Francesco and Perez Lozano, Daniel and Bengtsson, Andreas and Scarlino, Pasquale and Wallraff, Andreas and Chang, Darrick and Delsing, Per and Gasparinetti, Simone},
  journal = {Phys. Rev. X},
  volume = {12},
  issue = {3},
  pages = {031036},
  numpages = {22},
  year = {2022},
  month = {Sep},
  publisher = {American Physical Society},
  doi = {10.1103/PhysRevX.12.031036},
  url = {https://link.aps.org/doi/10.1103/PhysRevX.12.031036}
}

@misc{hernandez-anton_emission_2026,
	title = {Emission and {Absorption} of {Microwave} {Photons} in {Orthogonal} {Temporal} {Modes} across a 30-{Meter} {Two}-{Node} {Network}},
	copyright = {arXiv.org perpetual, non-exclusive license},
	url = {https://arxiv.org/abs/2604.12947},
	doi = {10.48550/ARXIV.2604.12947},
	urldate = {2026-04-16},
	publisher = {arXiv},
	author = {Hernández-Antón, Alonso and Schär, Josua D. and Grigorev, Aleksandr and Peñas, Guillermo F. and Puebla, Ricardo and García-Ripoll, Juan José and Besse, Jean-Claude and Wallraff, Andreas and Kulikov, Anatoly},
	year = {2026},
	note = {arXiv:2604.12947 [quant-ph]},
}

@article{almanakly_deterministic_2025,
	title = {Deterministic remote entanglement using a chiral quantum interconnect},
	volume = {21},
	issn = {1745-2473, 1745-2481},
	url = {https://www.nature.com/articles/s41567-025-02811-1},
	doi = {10.1038/s41567-025-02811-1},
	number = {5},
	urldate = {2026-05-04},
	journal = {Nat. Phys.},
	author = {Almanakly, Aziza and Yankelevich, Beatriz and Hays, Max and Kannan, Bharath and Assouly, Réouven and Greene, Alex and Gingras, Michael and Niedzielski, Bethany M. and Stickler, Hannah and Schwartz, Mollie E. and Serniak, Kyle and Wang, Joel I-j. and Orlando, Terry P. and Gustavsson, Simon and Grover, Jeffrey A. and Oliver, William D.},
	month = may,
	year = {2025},
	pages = {825--830},
}

@article{kannan_-demand_2023,
	title = {On-demand directional microwave photon emission using waveguide quantum electrodynamics},
	volume = {19},
	issn = {1745-2473, 1745-2481},
	url = {https://www.nature.com/articles/s41567-022-01869-5},
	doi = {10.1038/s41567-022-01869-5},
	number = {3},
	urldate = {2026-05-04},
	journal = {Nat. Phys.},
	author = {Kannan, Bharath and Almanakly, Aziza and Sung, Youngkyu and Di Paolo, Agustin and Rower, David A. and Braumüller, Jochen and Melville, Alexander and Niedzielski, Bethany M. and Karamlou, Amir and Serniak, Kyle and Vepsäläinen, Antti and Schwartz, Mollie E. and Yoder, Jonilyn L. and Winik, Roni and Wang, Joel I-Jan and Orlando, Terry P. and Gustavsson, Simon and Grover, Jeffrey A. and Oliver, William D.},
	month = mar,
	year = {2023},
	pages = {394--400},
}

@misc{luo_dynamic_2025,
	title = {Dynamic stimulated emission for deterministic addition and subtraction of propagating photons},
	url = {http://arxiv.org/abs/2512.09711},
	doi = {10.48550/arXiv.2512.09711},
	urldate = {2026-05-04},
	publisher = {arXiv},
	author = {Luo, Haoyuan and Shah, Parth S. and Yang, Frank and Mirhosseini, Mohammad and Mahmoodian, Sahand},
	month = dec,
	year = {2025},
	note = {arXiv:2512.09711 [quant-ph]},
}

@article{levy-yeyati_passive_2025,
	title = {Passive {Photonic} {CZ} {Gate} with {Two}-{Level} {Emitters} in {Chiral} {Multimode} {Waveguide} {QED}},
	volume = {6},
	issn = {2691-3399},
	url = {https://link.aps.org/doi/10.1103/PRXQuantum.6.010342},
	doi = {10.1103/PRXQuantum.6.010342},
	number = {1},
	urldate = {2026-05-04},
	journal = {PRX Quantum},
	author = {Levy-Yeyati, Tomás and Vega, Carlos and Ramos, Tomás and González-Tudela, Alejandro},
	month = mar,
	year = {2025},
	pages = {010342},
}

@article{kuang_perfect_2025,
	title = {Passive Quantum State Transfer in a Dispersion-Engineered Waveguide},
  author = {Kuang, Zeyu and Diekmann, Oliver and Fischer, Lorenz and Rotter, Stefan and Gonzalez-Ballestero, Carlos},
  journal = {Phys. Rev. Lett.},
  volume = {137},
  issue = {10},
  pages = {103605},
  numpages = {10},
  year = {2026},
  month = Sep,
  publisher = {American Physical Society},
  doi = {10.1103/m2md-rxkv},
  url = {https://link.aps.org/doi/10.1103/m2md-rxkv}
}

@article{du_dressed_2025,
	title = {Dressed {Interference} in {Giant} {Superatoms}: {Entanglement} {Generation} and {Transfer}},
	volume = {135},
	issn = {0031-9007, 1079-7114},
	shorttitle = {Dressed {Interference} in {Giant} {Superatoms}},
	url = {https://link.aps.org/doi/10.1103/crzs-k718},
	doi = {10.1103/crzs-k718},
	number = {22},
	urldate = {2026-05-08},
	journal = {Phys. Rev. Lett.},
	author = {Du, Lei and Wang, Xin and Kockum, Anton Frisk and Splettstoesser, Janine},
	month = nov,
	year = {2025},
	pages = {223601},
}
\end{document}